\documentclass[12pt, a4paper]{article}

 \usepackage[a4paper, top=2.0cm, bottom=1.0cm, left=1.5cm, right=1.5cm, footskip = 1.0cm, includehead, includefoot]{geometry}
 
\usepackage{helvet}

\usepackage{csquotes}
\usepackage[utf8]{inputenc}
\usepackage[final]{graphicx}
\usepackage[margin=20mm, font=footnotesize, labelfont=bf, justification=justified]{caption}
\usepackage{dirtytalk}
\usepackage{authblk}
\usepackage{blindtext}
\usepackage{hyperref}

\usepackage{times}
\usepackage[bottom]{footmisc}	
\usepackage{scrextend}
\deffootnote[3.5em]{3.5em}{2em}{\textsuperscript{\thefootnotemark}}

\usepackage{amsmath, amssymb, amsthm, graphicx, color, bm, soul}
\usepackage{amsmath}
\usepackage{xcolor}			
\usepackage{caption}
\usepackage{fourier}
\usepackage{mathtools} 

\usepackage{titlesec} 
\titleformat{\paragraph}
{\normalfont\normalsize\bfseries}{\theparagraph}{1em}{}

\usepackage{physics}
\usepackage{braket} 

\usepackage{musicography}  

\usepackage{fancyhdr}
\usepackage{tikz} 
\usetikzlibrary{quantikz}

\title{Quantum Computer Music: Foundations and Initial Experiments}

\author{Eduardo R. Miranda\thanks{Corresponding Author: eduardo.miranda@plymouth.ac.uk}}
\author{Suchitra T. Basak\thanks{suchitra.basak@plymouth.ac.uk}}
\affil{ICCMR, University of Plymouth, UK, http://cmr.soc.plymouth.ac.uk/}

\date{}     
                
\begin{document}

\maketitle

\abstract{Quantum computing is a nascent technology, which is advancing rapidly. There is a long history of research into using computers for music. Nowadays computers are absolutely essential for the music economy. Thus, it is very likely that quantum computers will impact the music industry in time to come. This chapter lays the foundations of the new field of \textit{Quantum Computer Music}. It begins with an introduction to algorithmic computer music and methods to program computers to generate music, such as Markov chains and random walks. Then, it presents  quantum computing versions of those methods. The discussions are supported by detailed explanations of quantum computing concepts and walk-through examples. A bespoke generative music algorithm is presented, the \textit{Basak-Miranda algorithm}, which leverages a property of quantum mechanics known as \textit{constructive and destructive interference} to operate a musical Markov chain. An Appendix introducing the fundamentals of quantum computing deemed necessary to understand the chapter and a link to access Jupyter Notebooks with examples are also provided.}

\section{Introduction}

\medskip
As early as the 1840s, mathematician - and most probably the first ever software programmer - Lady Ada Lovelace, predicted in that computers would be able to compose music. On a note about Charles Babbage’s Analytical Engine, she wrote:

\medskip
“\textit{Supposing, for instance, that the fundamental relations of pitched sounds in the science of harmony and of musical composition were susceptible of such expression and adaptations, the Engine might compose elaborate and scientific pieces of music of any degree of complexity or extent.}” (\cite{Manabrea1843}, p.21)

\medskip
At about the same time, steam powered machines controlled by stacks of punched cards were being engineered for the textile industry. Musical instrument builders promptly recognised that punch-card stacks could be used to drive automatic pipe organs. Such initiatives revealed a glimpse of an unsettling idea about the nature of the music they produced: it emanated from information, which could also be used to control all sorts of machines. The idea soon evolved into mechanical pianos (popularly known as ‘pianolas’) and several companies began as early as the 1900s to manufacture the so-called reproducing pianos. Reproducing pianos enabled pianists to record their work with good fidelity: the recording apparatus could punch thousands of holes per minute on a piano roll, enough to store all the notes that a fast virtuoso could play. Because a piano roll stored a set of parameters that represented musical notes rather than sound recordings, the performances remained malleable: the information could be manually edited, the holes re-cut, and so on. This sort of information technology gained much sophistication during the course of the twentieth century, and paved the way for the development of programmable electronic computers.   

\medskip
People do not often connect the dots to realise that field of Computer Music has been progressing in tandem with Computer Science since the invention of the computer. Musicians started experimenting with computing far before the emergence of the vast majority of scientific, industrial and commercial computing applications in existence today.

\medskip
For instance, as early as the 1940s, researchers at Australia’s Council for Scientific and Industrial Research (CSIR) installed a loudspeaker on their Mk1 computer to track the progress of a program using sound. Subsequently, Geoff Hill, a mathematician with a musical background, programmed this machine to playback a tune in 1951 \cite{Doornbusch2004}. Essentially, they programmed the Mk1 as if they were punching a piano roll for a pianola.

\begin{figure}[htbp]
\begin{center}\vspace{0.3cm}
\includegraphics[width=0.4\linewidth]{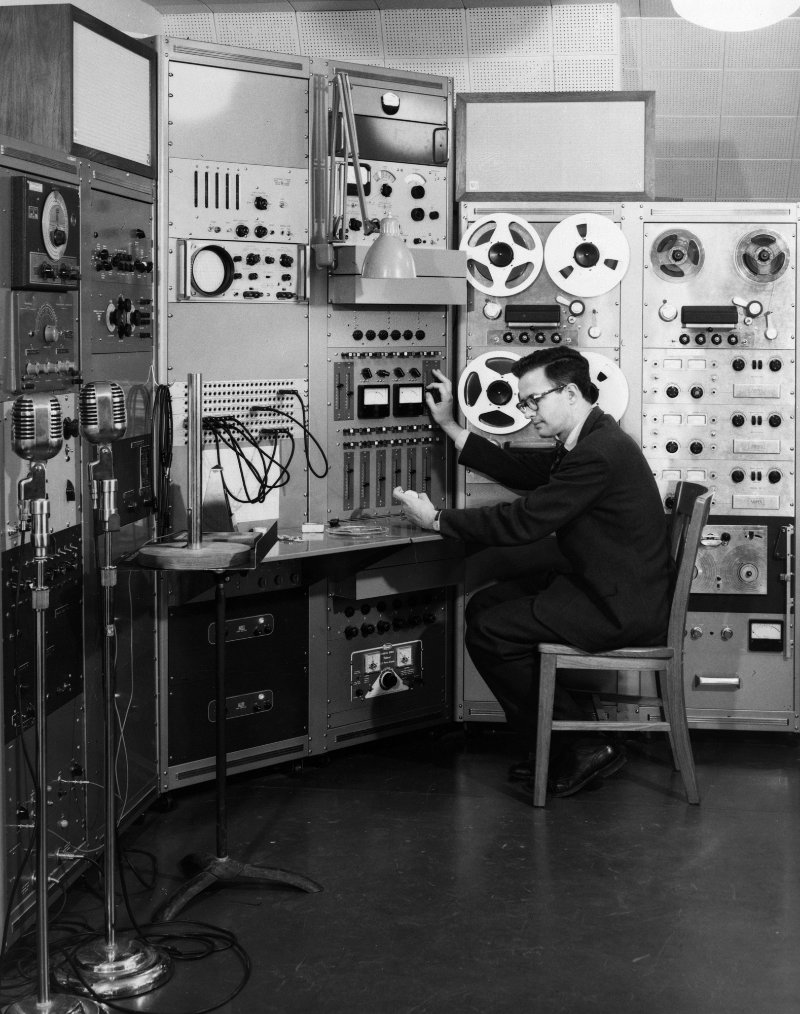}
\caption{Composer and Professor of Chemistry, Lejaren Hiller, in the University of Illinois at Urbana Champaign’s Experimental Music Studio. (Image courtesy of the University of Illinois at Urbana-Champaign.)}
\label{fig:hiller}
\end{center}
\end{figure}

\medskip
Then, in the 1950s Lejaren Hiller and Leonard Isaacson, at University of Illinois at Urbana-Champaign, USA, programmed the ILLIAC computer to compose a string quartet entitled \textit{Illiac Suite} (Figure \ref{fig:hiller}). The innovation here was that the computer was programmed with instructions to create music rather than merely reproduce encoded music. 

\medskip
The ILLIAC, short for Illinois Automatic Computer, was one of the first mainframe computers built in the Unites States, comprising thousands of vacuum tubes. The \textit{Illiac Suite} consists of four movements, each of which using different methods for generating musical sequences, including hard-coded rules and a probabilistic Markov chain method \cite{HillerIsaacson1959}. 

\medskip
The first uses of computers in music were for composition. The great majority of computer music pioneers were composers interested in inventing new music and/or innovative approaches to compose. They focused on developing algorithms to generate music. Hence the term ‘algorithmic computer music’. Essentially, the art of algorithmic computer music consists of (a) harnessing algorithms to produce patterns of data and (b) developing ways to translate these patterns into musical notes or synthesised sound. 

\medskip
Nowadays, computing technology is omnipresent in almost every aspect of music. Therefore, forthcoming alternative computing technology, such as biocomputing and quantum computing will certainly have an impact in the way in which we create and distribute music in time to come.

\medskip
This chapter introduces pioneering research into exploring emerging quantum computing technology in music. We say \enquote{emerging} because quantum computers are still being developed as we write this. There is some hardware already available, even commercially. However, detractors say that meaningful quantum computers, that is, quantum machines that can outperform current classical ones, are yet to be seen. Nevertheless, research and development is progressing fast. 

\medskip 
The chapter begins with an introduction to algorithmic computer music and methods to program computers to compose music, such as Markov chains and random walks \cite{Miranda2001}. Then, it discusses how to implement quantum computing versions of those methods. For didactic purposes, the discussions are supported by detailed explanations of basic quantum computing concepts and practical examples. A novel generative music algorithm in presented, which leverages a property of quantum mechanics known as \textit{constructive and destructive interference} 
\cite{Bernhardt2019} to operate Markov chains \cite{Privault2013} representing rules for sequencing music. An Appendix introducing the fundamentals of quantum computing deemed necessary to understand the chapter is also provided. Jupyter Notebooks with coding examples are available at the \href{https://iccmr-quantum.github.io/}{QuTune Project Website}\footnote{https://iccmr-quantum.github.io/}.

%
%

\section{Algorithmic Computer Music}

\medskip
An early approach to algorithmic computer music, which still remains popular to date, is to program a machine with rules for generating sequences of notes. Rules derived from classic treatises on musical composition are relatively straightforward to encode in a piece of software. Indeed, one of the movements of the \textit{Illiac Suite} string quartet mentioned earlier was generated with rules for making musical counterpoint\footnote{{\footnotesize In music, counterpoint is the art of combining different melodic lines in parallel.}} from a well-known treatise entitled \textit{Gradus ad Parnassum}, penned by Joseph Fux in 1725 \cite{Mann1965}.

\medskip
Rules for musical composition can be represented in a number of ways, including graphs, set algebra, Boolean expressions, finite state automata and Markov chains, to cite but five. For an introduction to various representation schemes and algorithmic composition methods please refer to the book \textit{Composing Music with Computers} \cite{Miranda2001}.

\medskip
As an example, consider the following 12-tone series derived from a chromatic scale starting with the pitch E: \{E, F, G, C\musSharp{}, F\musSharp{}, D\musSharp{}, G\musSharp{}, D, B, C, A, A\musSharp{}\}. Unless stated otherwise, the musical examples presented in this chapter assume the notion of 'pitch classes'. A pitch class encompasses all pitches that are related by octave or enharmonic equivalence. For instance, pitch class G\musSharp{} can be on any octave. Moreover, it sounds identical to pitch class A\musFlat{}; e.g., think of the a piano keyboard where the same black key plays G\musSharp{} and A\musFlat{}.

\begin{figure}[htbp]
\begin{center}\vspace{0.3cm}
\includegraphics[width=0.5\linewidth]{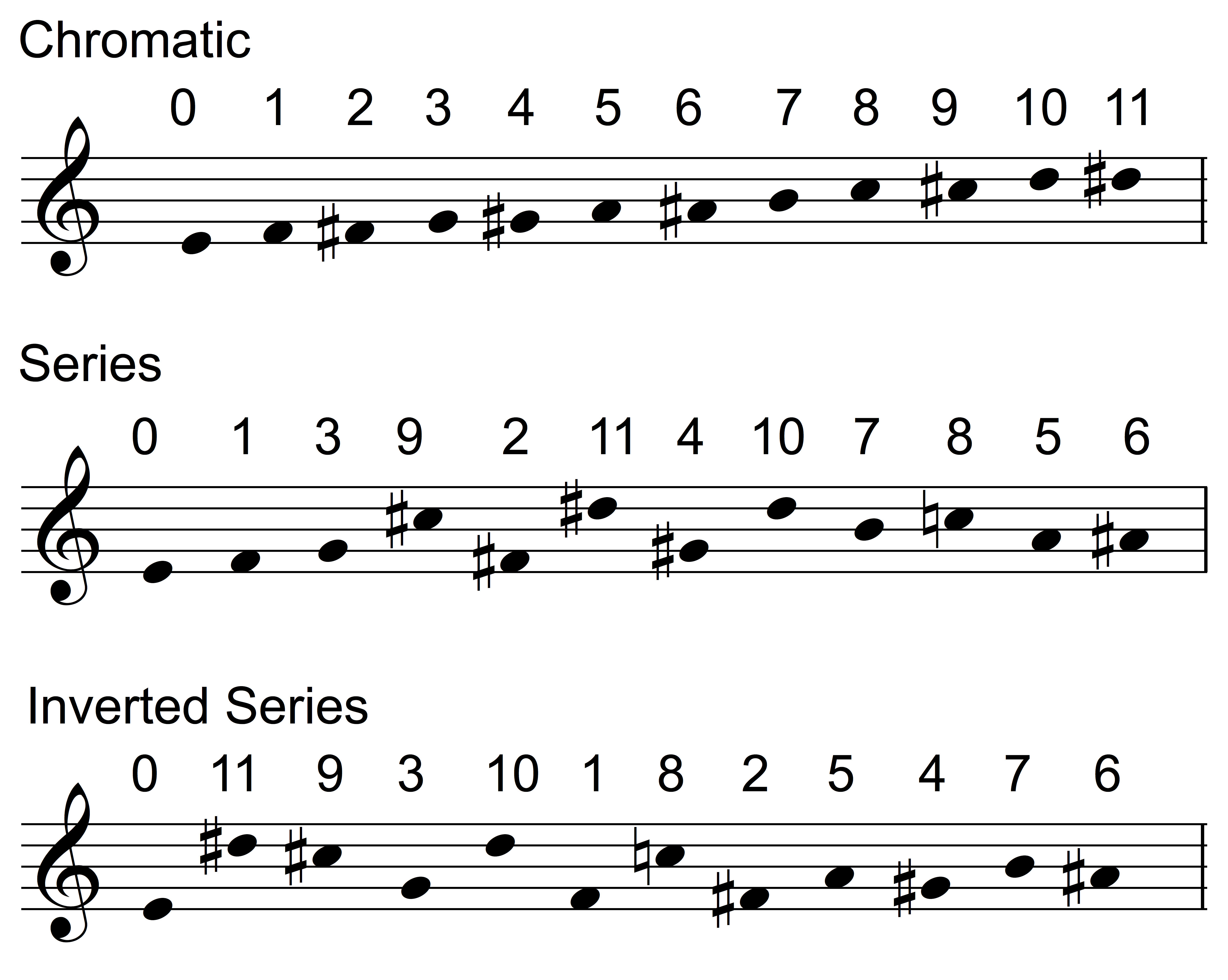}
\caption{Deriving a 12-tone series and its inverted version from a chromatic scale.}
\label{fig:series}
\end{center}
\end{figure}

\medskip
Figure \ref{fig:series} shows our 12-tone series in musical notation, its chromatic scale of origin and its inverted version, which will be used in the example that follows. The inverted version of a series is produced by reversing the intervals between the notes in the sequence; that is, going in the opposite direction of the original.  For example, a rising minor third (C $\rightarrow$ E\musFlat{}) becomes a falling minor third (C $\rightarrow$ A). For clarity, the numbers above each note in Figure \ref{fig:series} indicate its position in the chromatic scale.

\medskip
Now, let us define a bunch of sequencing rules. These are represented in Table \ref{table:rulesblocks}. The 12-tone series is laid on the horizontal axis and its inverted version on the vertical one. The rules are inspired by a composition method referred to as \textit{serialism}, popularised in the second half of the 20\textsuperscript{th} century by composers such as Karlheinz Stockhausen and Pierre Boulez \cite{Perle1972}. The details of this method are not important to dicuss here. What is important to observe is that for each note of the inverted series (in the vertical axis), there is a rule establishing which notes of the 12-tone series (in the horizontal axis) can follow it. For instance, the first row states that only an F or a D\musSharp{} can follow an E. 

\begin{table}[]
\centering
\begin{tabular}{|l|l|l|l|l|l|l|l|l|l|l|l|l|}
\hline
\multicolumn{1}{|c|}{}  & \multicolumn{1}{c|}{E} & \multicolumn{1}{c|}{F} & \multicolumn{1}{c|}{G} & \multicolumn{1}{c|}{C\musSharp{}} & F\musSharp{} & D\musSharp{} & G\musSharp{} & D & B  & C & A & A\musSharp{} \\ \hline
\multicolumn{1}{|c|}{E} & \multicolumn{1}{c|}{}  & \multicolumn{1}{c|}{$\blacksquare$} & \multicolumn{1}{c|}{}  & \multicolumn{1}{c|}{}  &   & $\blacksquare$ &   &   &   &   &   &   \\ \hline
\multicolumn{1}{|c|}{D\musSharp{}} & \multicolumn{1}{c|}{$\blacksquare$} & \multicolumn{1}{c|}{}  & \multicolumn{1}{c|}{}  & \multicolumn{1}{c|}{$\blacksquare$} & $\blacksquare$ &   & $\blacksquare$ &   &   &   &   &   \\ \hline
\multicolumn{1}{|c|}{C\musSharp{}} & \multicolumn{1}{c|}{}  & \multicolumn{1}{c|}{}  & \multicolumn{1}{c|}{$\blacksquare$} & \multicolumn{1}{c|}{}  & $\blacksquare$ & $\blacksquare$ &   &   &   &   &   &   \\ \hline
G & & $\blacksquare$ & & $\blacksquare$ & & & & $\blacksquare$ & & & &   \\ \hline
D & & $\blacksquare$ & $\blacksquare$ & & & & $\blacksquare$ & & $\blacksquare$ & & &  \\ \hline
F & $\blacksquare$ & & $\blacksquare$ & & & & & $\blacksquare$ &  & $\blacksquare$ & &   \\ \hline
C & & $\blacksquare$ & & & $\blacksquare$ & & & & $\blacksquare$ & & $\blacksquare$ &   \\ \hline
F\musSharp{} & &  & & $\blacksquare$ &  & $\blacksquare$ & & & & $\blacksquare$ & $\blacksquare$ &   \\ \hline
A & & & &  & $\blacksquare$ & & $\blacksquare$ & & & $\blacksquare$ &  & $\blacksquare$ \\ \hline
G\musSharp{} & & & & & & $\blacksquare$ & & $\blacksquare$ & $\blacksquare$ & & $\blacksquare$ &   \\ \hline
B & & & & & &  & $\blacksquare$ & $\blacksquare$ &   & $\blacksquare$ &   & $\blacksquare$ \\ \hline
A\musSharp{} & & &  & & & & & & $\blacksquare$ & & $\blacksquare$ &   \\ \hline
\end{tabular}
\caption{\footnotesize{Visual representation of sequencing rules. Columns are the notes of the series and rows are notes of the inverted series.}}
\label{table:rulesblocks}
\end{table}

\medskip
The rules are formalised as follows (the symbol $\lor$ stands for \enquote{or}):

\begin{itemize}
\item Rule 1: if E $\Longrightarrow$ F $\lor$ D\musSharp{}
\item Rule 2: if D\musSharp{} $\Longrightarrow$ E $\lor$ C\musSharp{} $\lor$ F\musSharp{} $\lor$ G\musSharp{}
\item Rule 3: if C\musSharp{} $\Longrightarrow$ G $\lor$ F\musSharp{} $\lor$ D\musSharp{}
\item Rule 4: if G $\Longrightarrow$ F $\lor$ C\musSharp{} $\lor$ D
\item Rule 5: if D $\Longrightarrow$F $\lor$ G $\lor$ G\musSharp{} $\lor$ B
\item Rule 6: if F $\Longrightarrow$ E $\lor$ G $\lor$ D $\lor$C
\item Rule 7: if C $\Longrightarrow$ F $\lor$ F\musSharp{}, B $\lor$  A
\item Rule 8: if F\musSharp{} $\Longrightarrow$ C\musSharp{} $\lor$ D\musSharp{} $\lor$ C $\lor$ A
\item Rule 9: if A $\Longrightarrow$  F\musSharp{} $\lor$ G\musSharp{} $\lor$ C $\lor$ A\musSharp{}
\item Rule 10: if G\musSharp{}	$\Longrightarrow$ D\musSharp{} $\lor$ D $\lor$ B $\lor$A
\item Rule 11: if B $\Longrightarrow$ G\musSharp{} $\lor$ D $\lor$ C $\lor$ A\musSharp{}
\item Rule 12: if A\musSharp{} $\Longrightarrow$ B $\lor$ A
\end{itemize}

\medskip
One way to implement those rules in a piece of software is to program algorithms to produce notes according to probability distributions, which are equally weighted between the notes allowed by a respective rule.  For instance, in Rule 2, each of the allowed 4 notes has a 25\% chance of occurring after D\musSharp{}. Thus, we can re-write Table \ref{table:rulesblocks} in terms of such probability distributions. This forms a Markov chain (Figure  \ref{table:rulesMarkov}). 

\begin{table}[]
\centering
\begin{tabular}{|l|l|l|l|l|l|l|l|l|l|l|l|l|}
\hline
\multicolumn{1}{|c|}{}  & \multicolumn{1}{c|}{E} & \multicolumn{1}{c|}{F} & \multicolumn{1}{c|}{G} & \multicolumn{1}{c|}{C\musSharp{}} & F\musSharp{} & D\musSharp{} & G\musSharp{} & D & B  & C & A & A\musSharp{} \\ \hline
\multicolumn{1}{|c|}{E} & \multicolumn{1}{c|}{}  & \multicolumn{1}{c|}{{\footnotesize0.5}} & \multicolumn{1}{c|}{}  & \multicolumn{1}{c|}{}  &   & {\footnotesize 0.5 }&   &   &   &   &   &   \\ \hline
\multicolumn{1}{|c|}{D\musSharp{}} & \multicolumn{1}{c|}{{\footnotesize 0.25}} & \multicolumn{1}{c|}{}  & \multicolumn{1}{c|}{}  & \multicolumn{1}{c|}{\footnotesize{0.25}} & {\footnotesize 0.25} &   & {\footnotesize 0.25} &   &   &   &   &   \\ \hline
\multicolumn{1}{|c|}{C\musSharp{}} & \multicolumn{1}{c|}{}  & \multicolumn{1}{c|}{}  & \multicolumn{1}{c|}{{\footnotesize 0.33}} & \multicolumn{1}{c|}{}  & {\footnotesize 0.33} & {\footnotesize 0.33} &   &   &   &   &   &   \\ \hline
G & & {\footnotesize 0.33} & & {\footnotesize 0.33} & & & & {\footnotesize 0.33} & & & &   \\ \hline
D & & {\footnotesize 0.25} & {\footnotesize 0.25} & & & & {\footnotesize 0.25} & & {\footnotesize 0.25} & & &  \\ \hline
F & {\footnotesize 0.25} & & {\footnotesize 0.25} & & & & & {\footnotesize 0.25} &  & {\footnotesize 0.25} & &   \\ \hline
C & & {\footnotesize 0.25} & & & {\footnotesize 0.25} & & & & {\footnotesize 0.25} & & {\footnotesize 0.25} &   \\ \hline
F\musSharp{} & &  & & {\footnotesize 0.25} &  & {\footnotesize 0.25} & & & & {\footnotesize 0.25} & {\footnotesize 0.25} &   \\ \hline
A & & & &  & {\footnotesize 0.25} & & {\footnotesize 0.25} & & & {\footnotesize 0.25} &  & {\footnotesize 0.25} \\ \hline
G\musSharp{} & & & & & & {\small 0.25} & & {\small 0.25} & {\small 0.25} & & {\small 0.25} &   \\ \hline
B & & & & & &  & {\footnotesize 0.25} & {\footnotesize 0.25} & & {\footnotesize 0.25} & & {\footnotesize0.25} \\ \hline
A\musSharp{} & & &  & & & & & & {\footnotesize 0.5} & & {\footnotesize 0.5} &   \\ \hline
\end{tabular}
\caption{\footnotesize{Sequencing rules represented as a Markov chain.}}
\label{table:rulesMarkov}
\end{table}

\medskip
Markov chains are conditional probability systems where the likelihood of future events depends on one or more past events. The number of past events that are taken into consideration at each stage is known as the order of the chain. A Markov chain that takes only one predecessor into account is of first order. A  chain that considers the predecessor and the predecessor’s predecessor, is of second order, and so on. We will come back to our 12-tone Markov chain later. 

\medskip
For now, let us have a look at how to generate music with a method known as \textit{random walk}. Imagine a system that is programmed to play a musical instrument with 12 notes, organised according to our 12-tone series. This system is programmed in such a way that it can play notes up and down the instrument by stepping only one note at a time. That is, only the next neighbour on either side of the last played note can be played next. If the system has a probability $p$ to hit the note on the left side of the last played note, then it will have the probability $q =  1 - p$ to hit the one on the right. This is an example of a simple one-dimensional random walk.

\begin{figure}[htbp]
\begin{center}\vspace{0.3cm}
\includegraphics[width=0.9\linewidth]{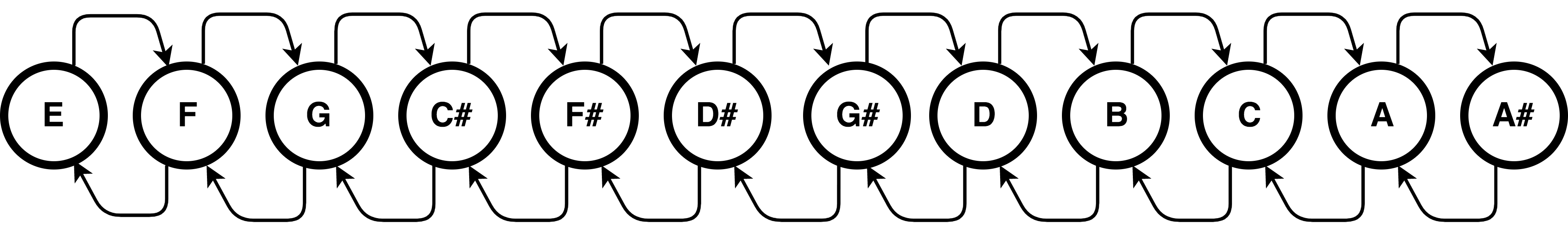}
\caption{Digraph representation of the simple random walk scheme depicted in Table \ref{table:randwalk}.}
\label{fig:digraphwalk}
\end{center}
\end{figure}

\medskip
A good way to visualize random walk processes is to depict them as directed graphs, or digraphs (Figure \ref{fig:digraphwalk}). Digraphs are diagram composed of nodes and arrows going from one node to another. Digraphs are widely used in computing to represent relationships in a network, such as roads linking places, hyperlinks connecting web pages, and so on.

\begin{table}[]
\centering
\begin{tabular}{|l|l|l|l|l|l|l|l|l|l|l|l|l|}
\hline
\multicolumn{1}{|c|}{}  & \multicolumn{1}{c|}{E} & \multicolumn{1}{c|}{F} & \multicolumn{1}{c|}{G} & \multicolumn{1}{c|}{C\musSharp{}} & F\musSharp{} & D\musSharp{} & G\musSharp{} & D & B  & C & A & A\musSharp{} \\ \hline
\multicolumn{1}{|c|}{E} & \multicolumn{1}{c|}{{\footnotesize \textbf{0.0}}} & \multicolumn{1}{c|}{{\footnotesize 1.0}} & \multicolumn{1}{c|}{} & \multicolumn{1}{c|}{} & & & & & & & &     \\ \hline
\multicolumn{1}{|c|}{F} & \multicolumn{1}{c|}{{\footnotesize 0.5}} & \multicolumn{1}{c|}{{\footnotesize \textbf{0.0}}} & \multicolumn{1}{c|}{{\footnotesize 0.5}} & \multicolumn{1}{c|}{} & & & & & & & &     \\ \hline
\multicolumn{1}{|c|}{G} & \multicolumn{1}{c|}{}    & \multicolumn{1}{c|}{{\footnotesize 0.5}} & \multicolumn{1}{c|}{{\footnotesize \textbf{0.0}}} & \multicolumn{1}{c|}{{\footnotesize 0.5}} & & & & & & & &     \\ \hline
C\musSharp{}  & & & {\footnotesize 0.5} & {\footnotesize \textbf{0.0}} & {\footnotesize 0.5} & & & & & & &     \\ \hline
F\musSharp{}  & & & & {\footnotesize 0.5} & {\footnotesize \textbf{0.0}} & {\footnotesize 0.5} & & & & & &     \\ \hline
D\musSharp{}  & & & & & {\footnotesize 0.5} & {\footnotesize \textbf{0.0}} & {\footnotesize 0.5} & & & & &     \\ \hline
G\musSharp{}  & & & & & & {\footnotesize 0.5} & {\footnotesize \textbf{0.0}} & {\footnotesize 0.5} & & & &     \\ \hline
D & & & & & & & {\footnotesize 0.5} & {\footnotesize \textbf{0.0}} & {\footnotesize 0.5} & & &     \\ \hline
B & & & & & & & & {\footnotesize 0.5} & {\footnotesize \textbf{0.0}} & {\footnotesize 0.5} & &     \\ \hline
C & & & & & & & & & {\footnotesize 0.5} & {\footnotesize\textbf{ 0.0}} & {\footnotesize 0.5} &     \\ \hline
A & & & & & & & & & & {\footnotesize 0.5} & {\footnotesize \textbf{0.0}} & {\footnotesize 0.5} \\ \hline
A\musSharp{} & & & & & & & & & & & {\footnotesize 1.0} & {\footnotesize \textbf{0.0}} \\ \hline
\end{tabular}
\caption{\footnotesize{Markov chain representation of a simple one-dimensional random walk process.(Empty cells are assumed to hold zeros.)}}
\label{table:randwalk}
\end{table}

\begin{figure}[htbp]
\begin{center}\vspace{0.6cm}
\includegraphics[width=0.5\linewidth]{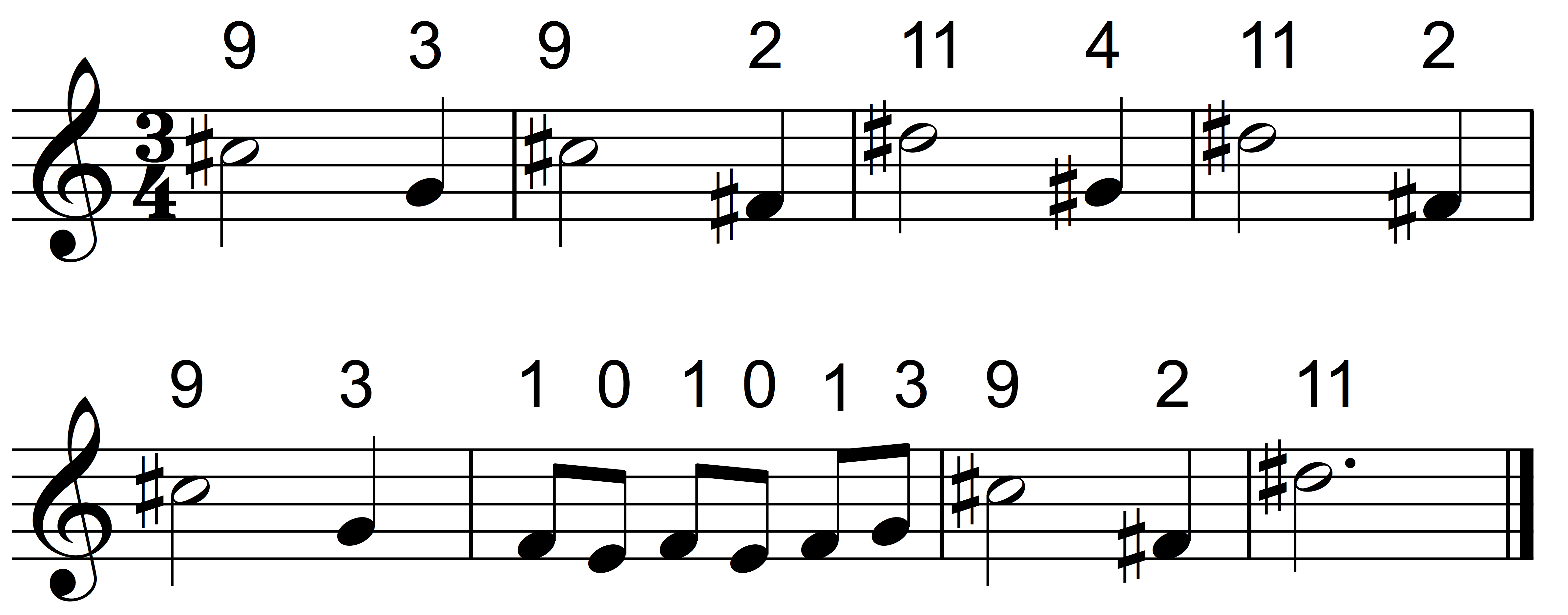}
\caption{A composition with pitches generated by the random walk depicted in Figure \ref{fig:digraphwalk}, using the series in Figure \ref{fig:series}. A number above each note indicates its position in the chromatic scale. Rhythmic figures were assigned manually.}
\label{fig:randwalk-composition}
\end{center}
\end{figure}

\medskip
As a matter of fact, the random walk depicted in Figure \ref{fig:digraphwalk} can be represented as a Markov chain with non-zero entries immediately on either side of the main diagonal, and zeros everywhere else. This is illustrated in Table \ref{table:randwalk}. Note that in this case we replaced the vertical axis of the previous tables with the same series as the one laid on the horizontal axis. Figure \ref{fig:randwalk-composition} shows an example of a sequence generated by our imaginary system. (Rhythmic figures were assigned manually.) Starting with pitch C\musSharp{}, a virtual die decided which one to pick next: G or F\musSharp{}. For this example, G was selected. Then, from G there were two options, and C\musSharp{} was selected.  And so on and so forth.

%
%

\section{Musical Quantum Walks}

This subsection introduces two preliminary ideas for making music with quantum versions of random walk processes. A comprehensive introduction to the basics of quantum computing and gate operations is beyond the scope of this chapter. Please refer to the Appendix for the very basics of gate-based quantum computing and to \cite{Rieffel2011} for a more detailed introduction.

\subsection{One-dimensional musical quantum walk}

\begin{figure}[htbp]
\begin{center}\vspace{0.3cm}
\includegraphics[width=0.6\linewidth]{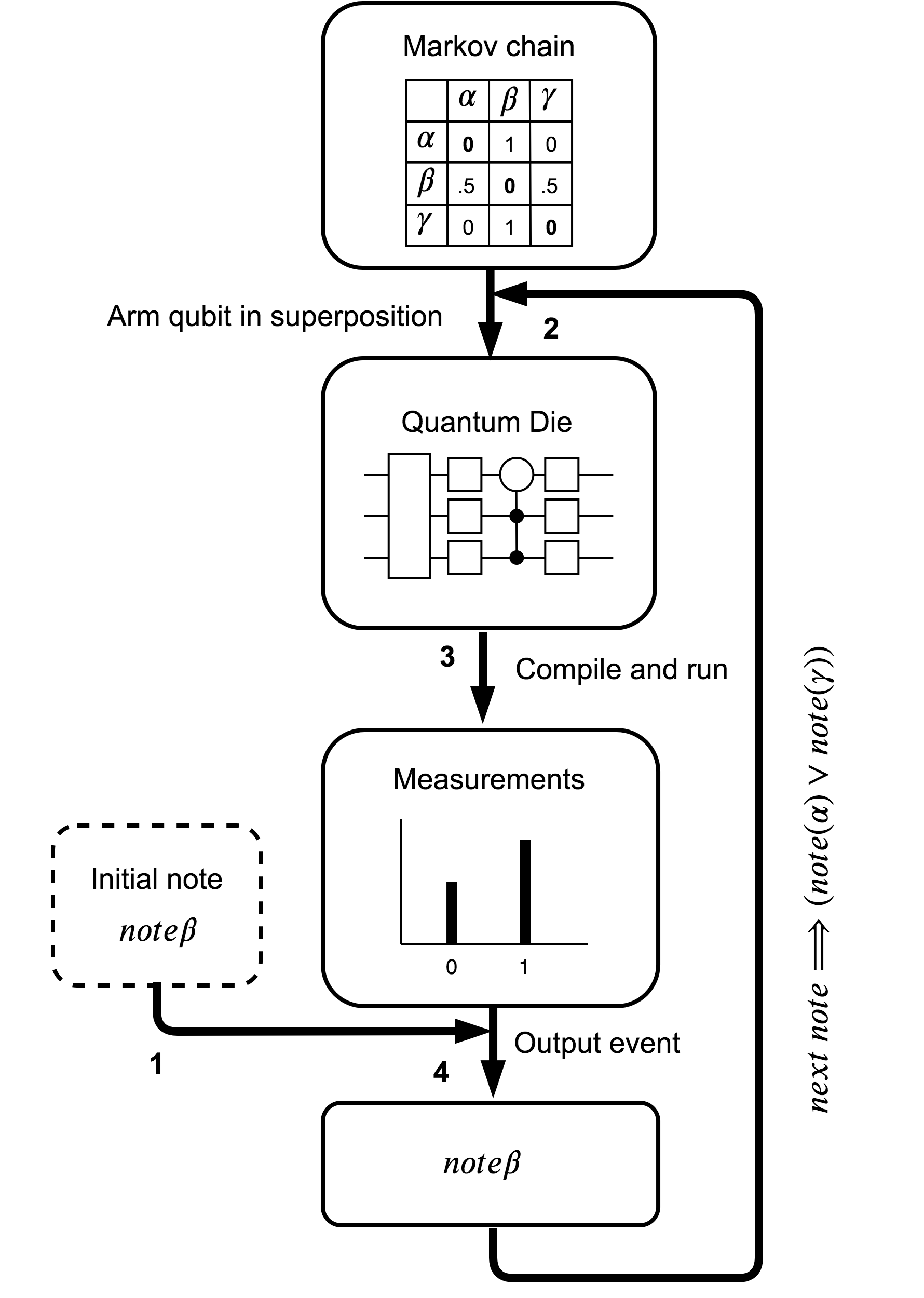}
\caption{One-dimensional quantum walk generative music system.}
\label{fig:1d-qwalk-circuit}
\end{center}
\end{figure}

\medskip
A quantum implementation of the one-dimensional quantum walk depicted in Table \ref{table:randwalk} is shown in Figure \ref{fig:1d-qwalk-circuit}.  

\medskip
Given an initial musical note from our 12-tone series (Figure \ref{fig:1d-qwalk-circuit} (1)), the system plays this note (4) and runs a binary quantum die (2, 3). The result is used to pick from the Markov chain the next note to play. If the result is equal to 0, then it picks the note on the left side of the last played note, otherwise it picks the one on the right side; see Figure \ref{fig:digraphwalk}. Next, the circuit is re-armed (2) and run again to pick another note. This process is repeated for a pre-specified number of times. 

\begin{figure}[htbp]
\begin{center}\vspace{0.3cm}
    \begin{tikzpicture}
        \node[scale=1.0] 
        {
            \begin{quantikz}
                \lstick{$q_0$} &  \ket{0} & \gate{\textbf{H}}  &  \meter{}  & \qw
             \end{quantikz}
        };
    \end{tikzpicture}
\end{center}
\caption{Circuit for a quantum die.}
\label{fig:die-circuit}
\end{figure}
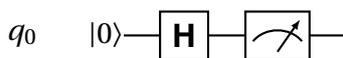

\begin{figure}[htbp]
\begin{center}\vspace{0.3cm}
\includegraphics[width=0.3\linewidth]{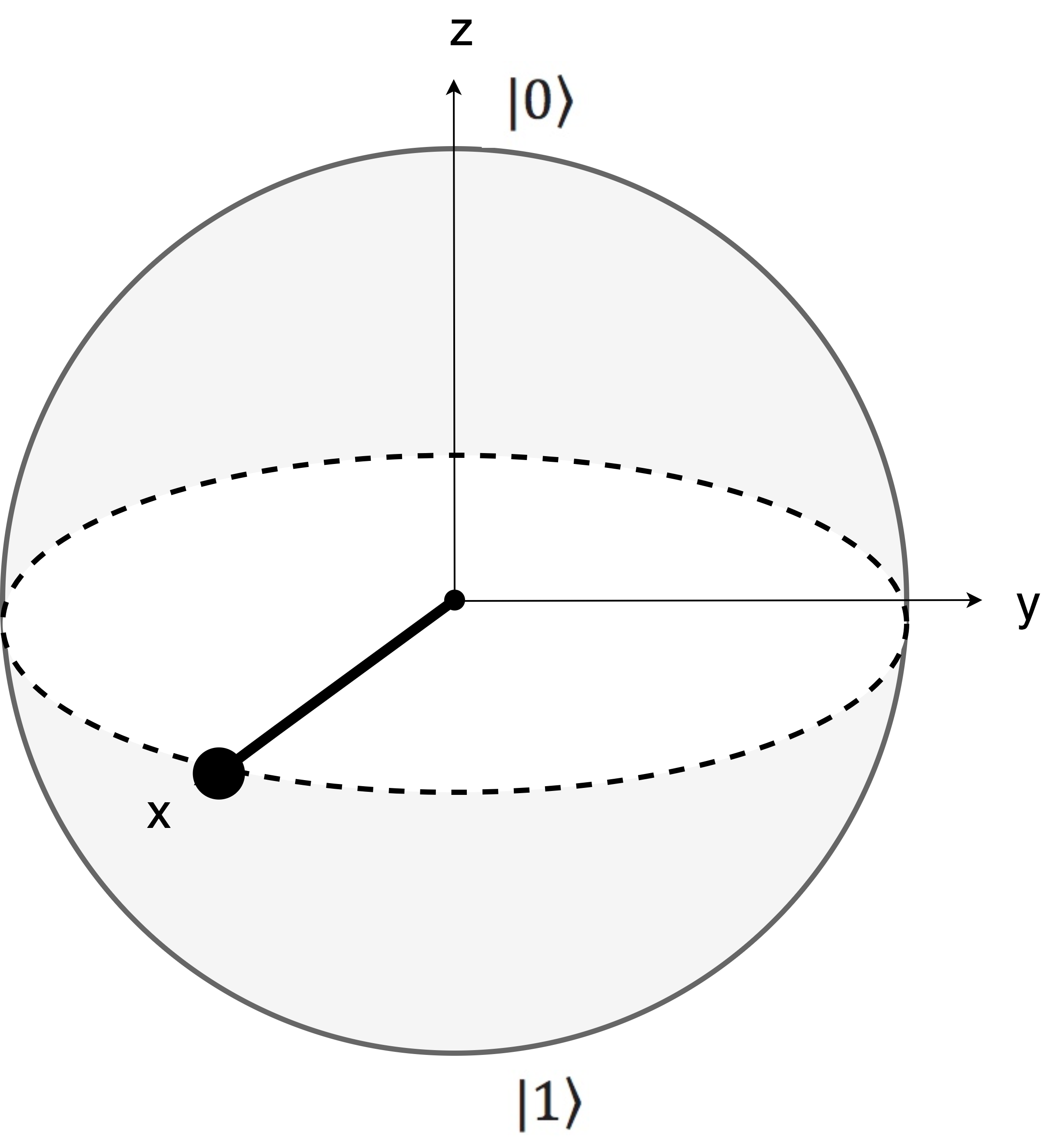}
\caption{Bloch sphere representation of a qubit in an equally-weighted state of superposition. The state vector is pointing at the equator line of the sphere. But when measured, the vector will move north or south, and will value, 0 or 1, respectively.}
\label{fig:Bloch-superposition}
\end{center}
\end{figure}

\medskip
The quantum binary die is implemented with a single qubit and a Hadamard, or H, gate (Figure \ref{fig:die-circuit}). The H gate puts the state vector $\ket{\Psi}$ of the qubit in an equally weighted combination of two opposing states, $\ket{0}$ and $\ket{1}$. This is represented on the Bloch sphere in Figure \ref{fig:Bloch-superposition}. Mathematically, this is notated as follows: $\ket{\Psi} = \alpha \ket{0} + \beta \ket{1}$ with $\left| \alpha \right|^2 = 0.5$ and $\left| \beta \right|^2 = 0.5$. That is to say, there is a $50\%$ chance of returning 0 or 1 when the qubit is measured.  

\subsection{Three-dimensional musical quantum walk}

\begin{figure}[htbp]
\begin{center}\vspace{0.3cm}
\includegraphics[width=0.4\linewidth]{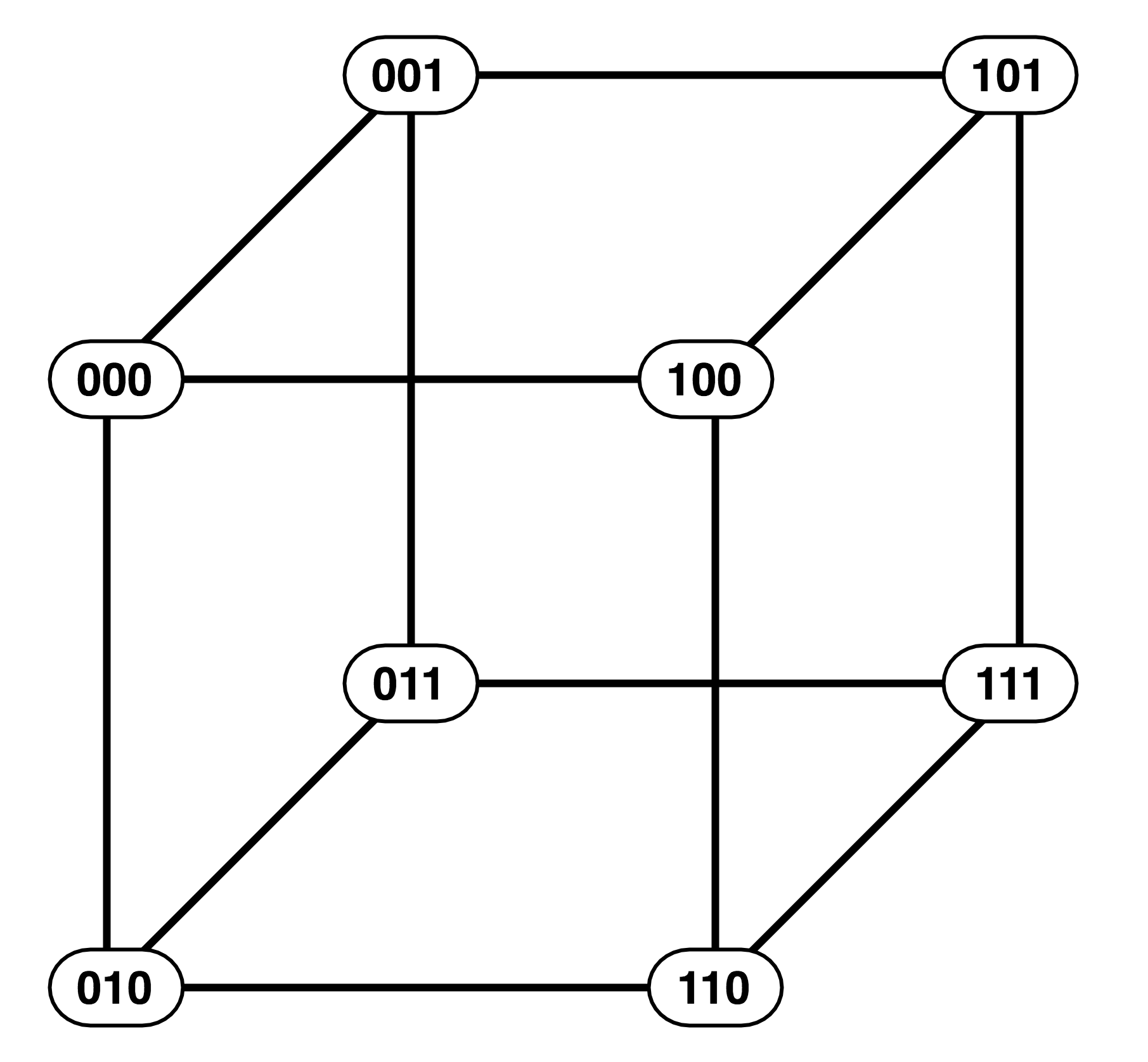}
\caption{Cube representation of the quantum walk.}
\label{fig:cube}
\end{center}
\end{figure}

\medskip
Let us take a look at a random walk scheme with more than two travelling options. As an example, consider that the vertices of the cube shown in Figure \ref{fig:cube} represent the nodes of a graph. And its edges represent the possible routes to move from one node to another. From, say, node 100 the walker could remain on 100, or go to 000, 110 or 101.

\medskip
In a classical random walk, a \enquote{walker} inhabits a definite node at any one moment in time. But in quantum walk, it would be in a superposition of all nodes it can possibly visit in a given moment. Metaphorically, we could say that the walker would stand on all viable nodes simultaneously, until it is observed, or measured. 

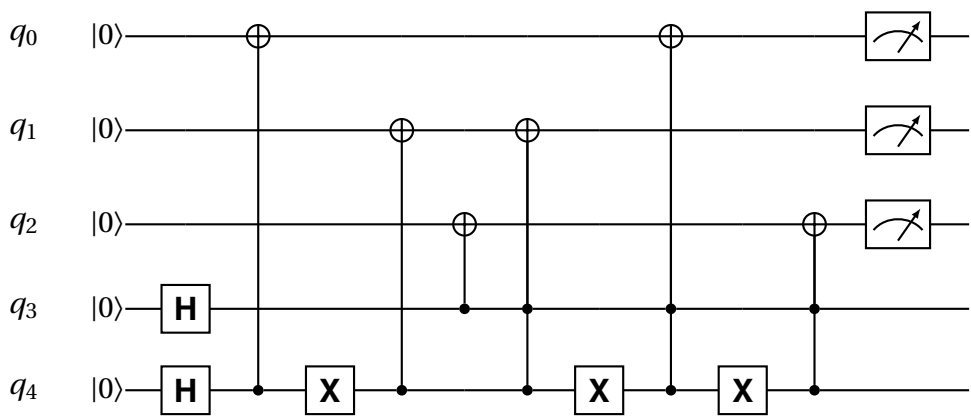
\begin{figure}[htbp]
\begin{center}\vspace{0.3cm}
    \begin{tikzpicture}
        \node[scale=1.0] 
        {
            \begin{quantikz}
                \lstick{$q_0$} & \ket{0} & \qw & \targ{} & \qw &  \qw{} & \qw & \qw & \qw & \targ{} & \qw  & \qw & \meter{} & \qw \\
                \lstick{$q_1$} & \ket{0} & \qw  & \qw & \qw &  \targ{} 	& \qw & \targ{} 	& \qw & \qw & \qw & \qw & \meter{} & \qw \\
                \lstick{$q_2$} & \ket{0} & \qw  & \qw & \qw  & \qw & \targ{} & \qw & \qw & \qw & \qw 	& \targ{} & \meter{} & \qw \\
                \lstick{$q_3$} & \ket{0} & \gate{\textbf{H}}  & \qw & \qw & \qw & \ctrl{-1} & \ctrl{-2} & \qw & \ctrl{-3}& \qw & \ctrl{-1}	& \qw & \qw \\
                \lstick{$q_4$} & \ket{0} & \gate{\textbf{H}}  & \ctrl{-4}  & \gate{\textbf{X}} & \ctrl{-3} & \qw & \ctrl{-3} & \gate{\textbf{X}} & \ctrl{-4} 	& \gate{\textbf{X}} & \ctrl{-2} & \qw & \qw
            \end{quantikz} 
        };
    \end{tikzpicture}
\end{center}
\caption{Quantum walk circuit.}
\label{fig:walk-circuit}
\end{figure}

\medskip
The quantum circuit in Figure \ref{fig:walk-circuit} moves a walker through the edges of the cube. The circuit uses five qubits: three ($q_0$, $q_1$, and $q_2$) to encode the eight vertices of the cube \{000, 001, …, 111\} and two ($q_3$ and $q_4$) to encode the possible routes that the walker can take from a given vertex, one of which is to stay put. 

\medskip
The circuit diagram shows a sequence of quantum operations, the first of which are two Hadamard gates applied to $q_3$ and $q_4$, followed by a controlled NOT gate with $q_4$ as a control to flip the state vector of $q_0$, and so on. We refer to the first three qubits as \textit{input qubits} and the last two as \textit{dice qubits}. The dice qubits act as controls for NOT gates to invert the state vector of input qubits. 

\medskip
For every vertex of the cube, the edges connect three neighbouring vertices whose codes differ by changing only one bit of the origin’s code. For instance, vertex 111 is connected to  011, 101 and 110. Therefore, upon measurement the system returns one of four possible outputs: 

\begin{itemize}
\item the original input with flipped $q_0$
\item the original input with flipped $q_1$
\item the original input with flipped $q_2$
\item the original input unchanged
\end{itemize}

\medskip
The quantum walk algorithm runs as follows: the input qubits are armed with the state representing a node of departure and the two dice qubits are armed in balanced superposition (Hadamard gate). Then, the input qubits are measured and the results are stored in a classical memory. This causes the whole system to decohere. Depending on the values yielded by the dice qubits, the conditional gates will change the state of the input qubits accordingly. Note that we measure and store only input qubits; the values of the dice can be lost. The result of the measurements is then used to arm the input qubits for the next step of the walk, and the cycle continues for a number of steps\footnote{\footnotesize{In fact, each step is run for thousands of times, or thousands of shots in quantum computing terminology. The reason for this will be clarified later. \label{foot:foot_shots}}}. The number of steps is established at the initial data preparation stage.

\medskip
As a trace table illustration, let us assume the following input: 001,  where $q_0$ is armed to $\ket{0}$,  $q_1$ to $\ket{0}$ and $q_2$ to $\ket{1}$. Upon measurement, let us suppose that the dice yielded $q_3 = 0$ and $q_4 = 1$. The second operation on the circuit diagram is a controlled NOT gate where $q_4$ acts as a conditional to invert $q_0$. Thus, at this point the state vector of $q_0$ is inverted because $q_4 = 1$. As the rest of the circuit does not incur any further action on the input qubits, the system returns 101. Should the dice have yielded $q_3 = 0$ and $q_4 = 0$ instead, then the third operation, which is a NOT gate, would have inverted $q_4$, which would subsequently act as a conditional to invert $q_1$. The result in this case would have been 011.

\begin{figure}[htbp]
\begin{center}\vspace{0.3cm}
\includegraphics[width=0.6\linewidth]{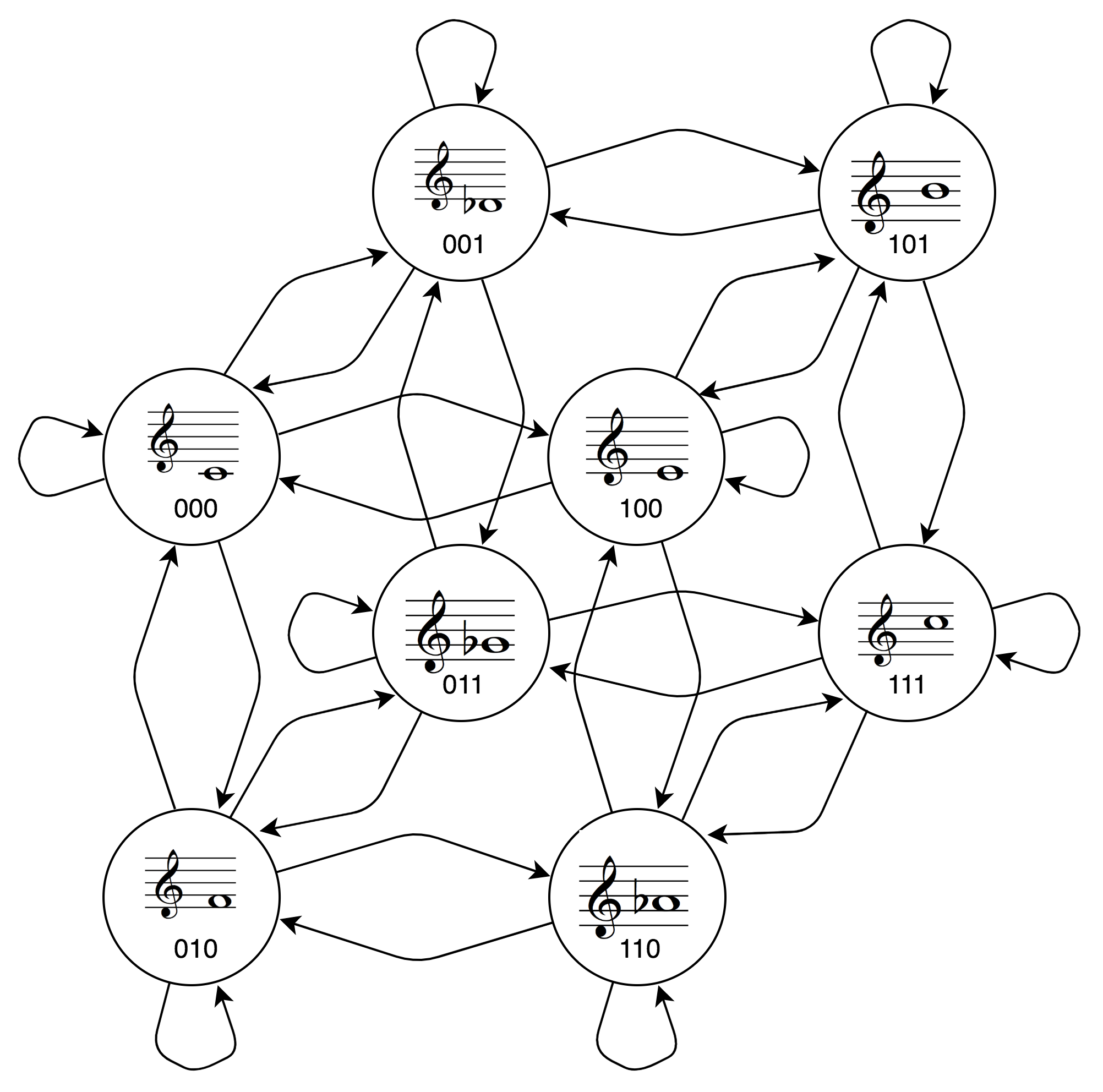}
\caption{Digraph representation of the quantum walk routes through musical notes. Notes from the so-called Persian scale on C are associated to the vertices of the cube shown in Figure \ref{fig:cube}.}
\label{fig:persian}
\end{center}
\end{figure}

\begin{figure}[htbp]
\begin{center}\vspace{0.3cm}
\includegraphics[width=0.4\linewidth]{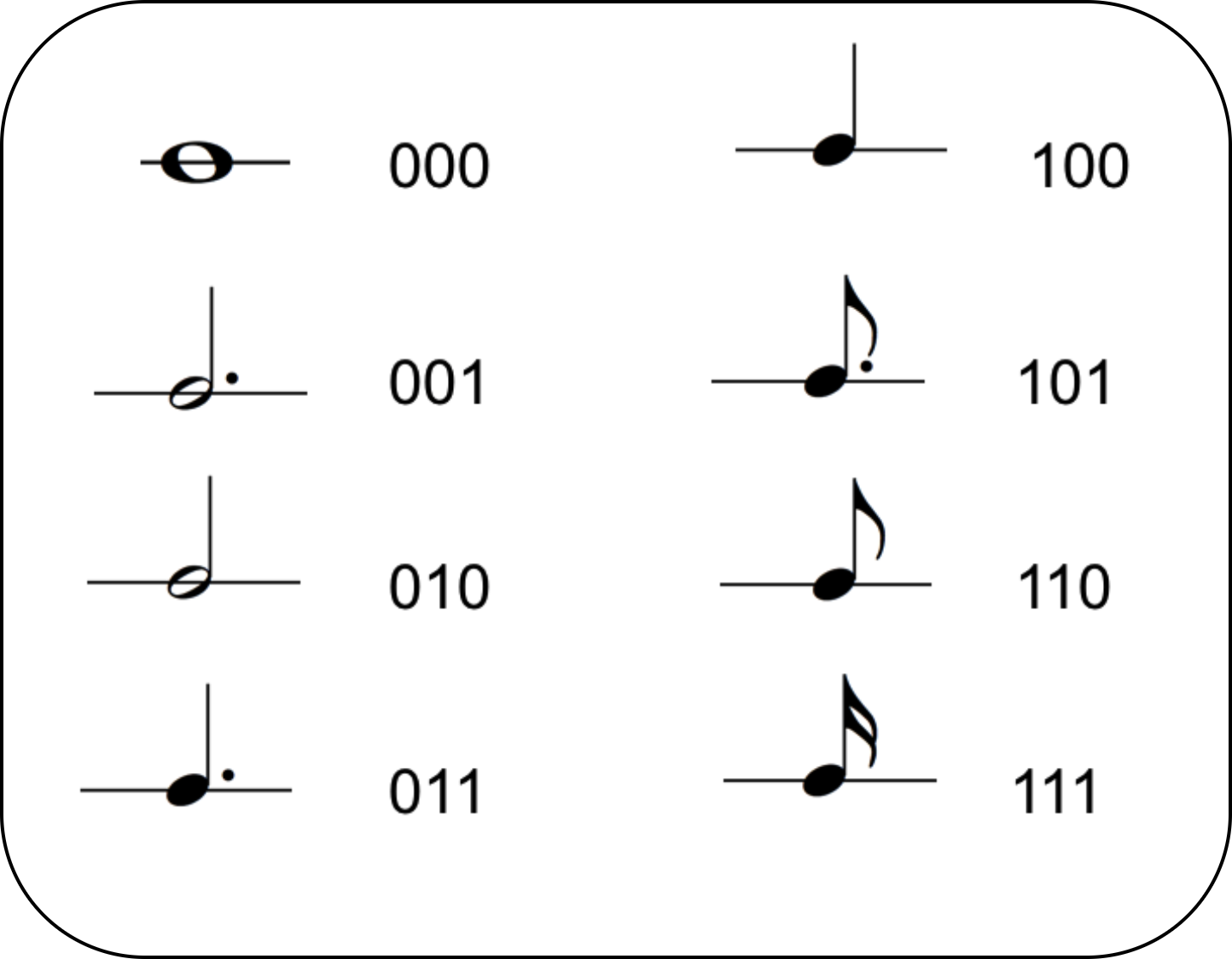}
\caption{Rhythmic figures associated with the notes of the cube shown in Figure \ref{fig:cube}.}
\label{fig:rhythmic-figs}
\end{center}
\end{figure}

\medskip
We implemented a demonstration system that runs each cycle of the quantum walk twice: once to generate a pitch and once again to generate a rhythmic figure. Take it that the system walks through two separate cubes in parallel. One of the cubes, say the \textit{pitch-cube}, encodes pitches on its vertices. The other, say the \textit{rhythm-cube}, encodes rhythms. Figure \ref{fig:persian} shows a digraph representation to the pitch-cube, where each vertex is associated with a pitch; the respective binary code is also shown. The eight pitches form a musical scale known as the Persian scale on C:  \{C4, D\musFlat{}4, E4, F4, G\musFlat{}4, A\musFlat{}4, B4, C5\}. Figure \ref{fig:rhythmic-figs} shows the rhythmic figures associated to the vertices of the rhythm-cube. The system holds different musical dictionaries associating vertices with different sets of pitches and rhythmic figures\footnote{The length of the composition and the dictionaries are set up by the use. It is possible to change dictionaries automatically while the system is generating the music.}

\begin{table}[htbp]
\centering
\begin{tabular}{|c|c|c|c|c|c|c|}
\hline
\textbf{{\footnotesize Step}} & \textbf{{\footnotesize Pitch}} & \textbf{{\footnotesize Rhythm}} &  & \textbf{{\footnotesize Step}} & \textbf{{\footnotesize Pitch}} & \textbf{{\footnotesize Rhythm}} \\ \hline
\textbf{{\footnotesize 0}}    & {\footnotesize 000} & {\footnotesize 100} & & \textbf{{\footnotesize 15}} & {\footnotesize 110} & {\footnotesize 100} \\ \hline
\textbf{{\footnotesize 1}}    & {\footnotesize 000} & {\footnotesize 000} & & \textbf{{\footnotesize 16}} & {\footnotesize 111} & {\footnotesize 110}  \\ \hline
\textbf{{\footnotesize 2}}    & {\footnotesize 001} & {\footnotesize 010} & & \textbf{{\footnotesize 17}} & {\footnotesize 111} & {\footnotesize 100}  \\ \hline
\textbf{{\footnotesize 3}}    & {\footnotesize 000} & {\footnotesize 110} & & \textbf{{\footnotesize 18}} & {\footnotesize 110} & {\footnotesize 101}  \\ \hline
\textbf{{\footnotesize 4}}    & {\footnotesize 000} & {\footnotesize 100} & & \textbf{{\footnotesize 19}} & {\footnotesize 110} & {\footnotesize 100}  \\ \hline
\textbf{{\footnotesize 5}}    & {\footnotesize 001} & {\footnotesize 101} & & \textbf{{\footnotesize 20}} & {\footnotesize 010} & {\footnotesize 100}  \\ \hline
\textbf{{\footnotesize 6}}    & {\footnotesize 001} & {\footnotesize 001} & & \textbf{{\footnotesize 21}} & {\footnotesize 011} & {\footnotesize 110}  \\ \hline
\textbf{{\footnotesize 7}}    & {\footnotesize 000} & {\footnotesize 001} & & \textbf{{\footnotesize 22}} & {\footnotesize 010} & {\footnotesize 010}  \\ \hline
\textbf{{\footnotesize 8}}    & {\footnotesize 001} & {\footnotesize 001} & & \textbf{{\footnotesize 23}} & {\footnotesize 010} & {\footnotesize 110}  \\ \hline
\textbf{{\footnotesize 9}}    & {\footnotesize 011} & {\footnotesize 101} & & \textbf{{\footnotesize 24}} & {\footnotesize 011} & {\footnotesize  010}  \\ \hline
\textbf{{\footnotesize 10}}  & {\footnotesize 011} & {\footnotesize 001} & & \textbf{{\footnotesize 25}} & {\footnotesize 011} & {\footnotesize 011}   \\ \hline
\textbf{{\footnotesize 11}}  & {\footnotesize 011} & {\footnotesize 000} & & \textbf{{\footnotesize 26}} & {\footnotesize 010} & {\footnotesize 010}  \\ \hline
\textbf{{\footnotesize 12}}  & {\footnotesize 001} & {\footnotesize 001} & & \textbf{{\footnotesize 27}} & {\footnotesize 000} & {\footnotesize 000}   \\ \hline
\textbf{{\footnotesize 13}}  & {\footnotesize 000} & {\footnotesize 001} & & \textbf{{\footnotesize 28}} & {\footnotesize 000} & {\footnotesize 000}   \\ \hline
\textbf{{\footnotesize 14}}  & {\footnotesize 010} & {\footnotesize 101} & & \textbf{{\footnotesize 29}} & {\footnotesize 000}  & {\footnotesize 010}   \\ \hline
\end{tabular}
\caption{\footnotesize{Results from running the quantum walk algorithm 30 times.}}
\label{tab:qwalk-results}
\end{table}

\begin{figure}[htbp]
\begin{center}\vspace{0.3cm}
\includegraphics[width=0.8\textwidth]{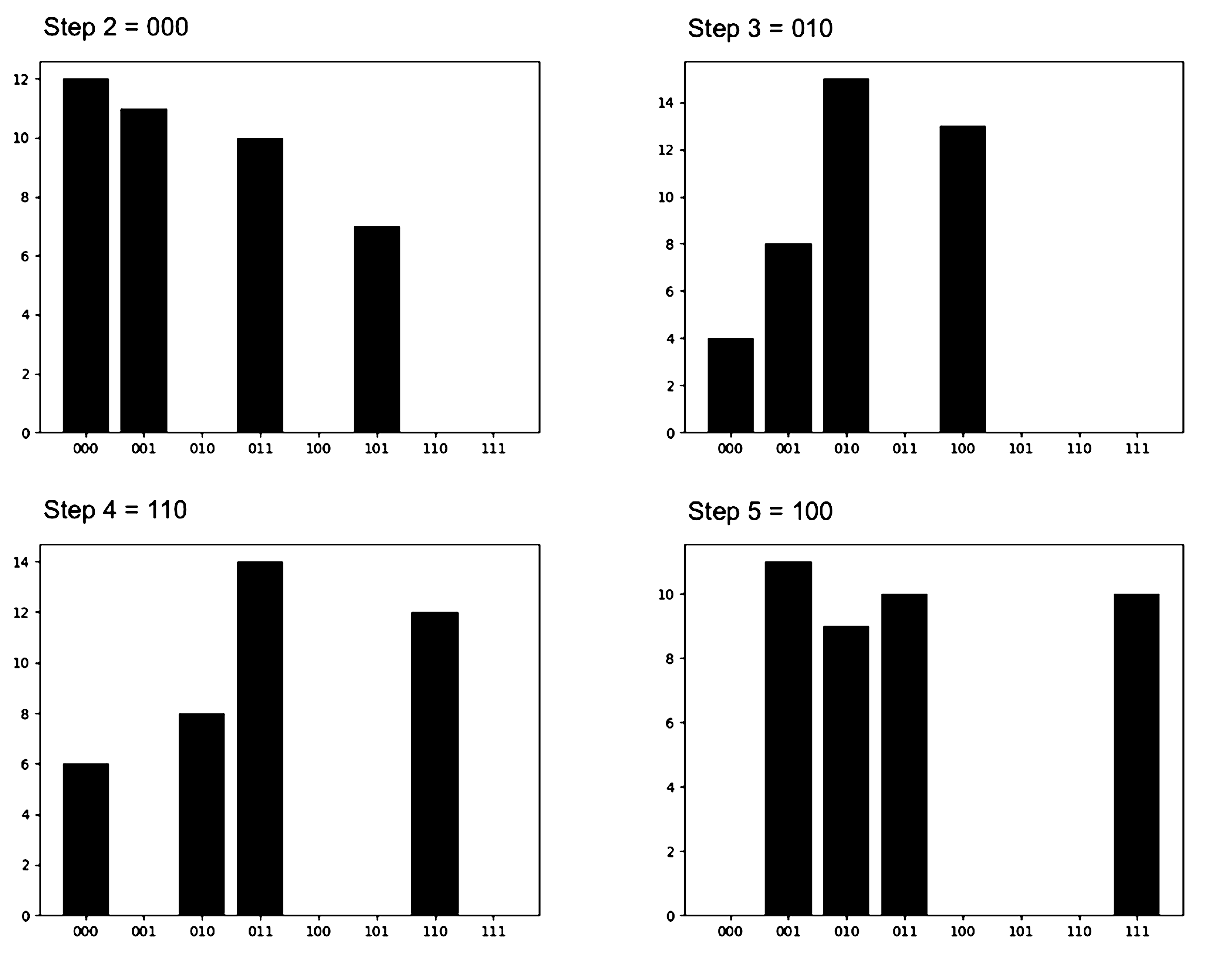}
\caption{Histograms for four steps of the quantum walk algorithm for generating rhythms. The vertical coordinate is the number of times an item listed on the horizontal coordinate occurred. For each step, the system selects the result that occurred more frequently. Note that the digits on the plots are in reverse order; this is due to the way in which the qubits are normally ordered in quantum programming.}
\label{fig:qwalk-histograms}
\end{center}
\end{figure}

\begin{figure}[htbp]
\begin{center}\vspace{0.3cm}
\includegraphics[width=0.6\linewidth]{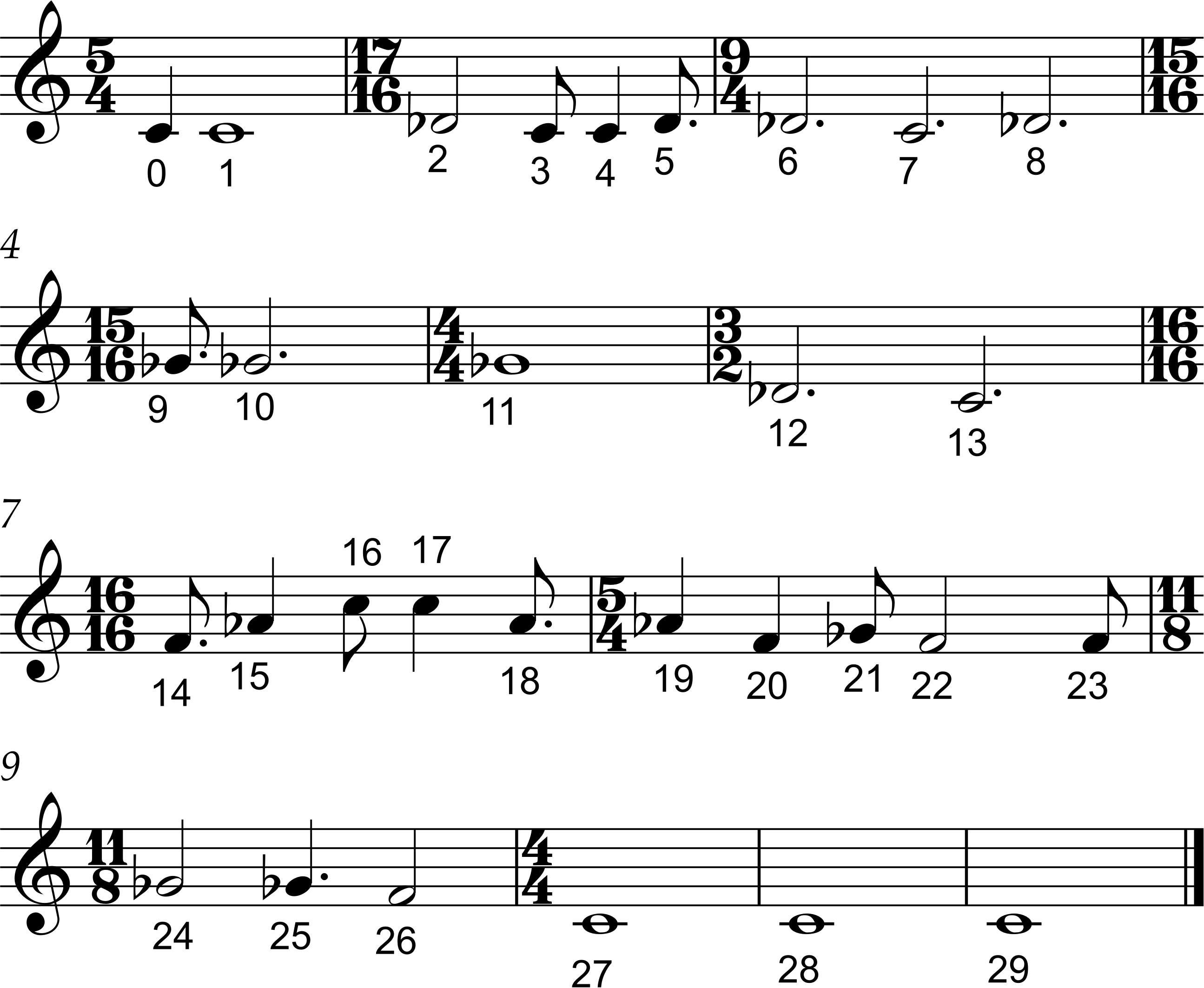}
\caption{A short composition generated by the quantum walk system.}
\label{fig:qwalk-composition}
\end{center}
\end{figure}
\medskip
To generate a musical sequence, the system starts with a given note\footnote{This initial note is input by the user.}; for instance, a crochet C4, whose codes for pitch and rhythm are 000 and 100, respectively. Then, for every new note the system runs the quantum walk circuit twice, once with input qubits armed with the code for pitch and then armed with the code for rhythm. The results from the measurements are then used to produce the next note. For instance, if the first run results in 000 and the second in 000, then the resulting note is another C4, but as a semibreve. The measurements are used to re-arm the circuit for the next note and so on.

\medskip
An example of a composition generated by the system is shown in Figure \ref{fig:qwalk-composition}. In this case the system ran for 29 steps. The initial pitch was C4 (code = 000) and the initial rhythmic figure was as crochet (code = 100). Below each note on the score in Figure \ref{fig:qwalk-composition} is a number indicating the step that generated it. Table \ref{tab:qwalk-results} shows the codes generated at each step of the walk. 

\medskip
As mentioned in footnote \ref{foot:foot_shots} due to the statistical nature of quantum computation, it is necessary to execute a quantum algorithm for multiple times, or shots, in order to obtain results that are statistically plausible. This enables one to inspect if the outcomes reflect the envisaged amplitude distribution of the quantum states. And running a circuit multiple times helps to mitigate the effect of errors. For each shot, the measurements are stored in standard digital memory, and in the case of our quantum walk algorithm, the result that occurred more frequently is selected.  Figure \ref{fig:qwalk-histograms} shows histograms for four steps, for 40 shots each, for generating rhythms. Starting on 100, then the walker moves to 000 in step 1, to 010 in step 2, to 110 in step 3, and 100 in step 4. As we generated the example on a quantum computing software simulator\footnote{\footnotesize{For this example we used Rigetti's Quantum Virtual Machine (QVM), implemented in ANSI Common LISP.}}, 40 shots for each step were sufficient to obtain the expected results.

\section{Basak-Miranda Algorithm}

\medskip
Throwing a one-qubit quantum die to select between two allowed choices on a Markov chain representing simple one-dimensional random walks works well. But this would not work for a chain representing more complex sequencing rules, such as the one in Table \ref{table:rulesMarkov}. 

\medskip
To tackle this, we designed a novel algorithm: the \textit{Basak-Miranda algorithm}. This algorithm exploits a fundamental property of quantum physics, known as \textit{constructive and destructive interference} \cite{Griffiths2018}, to select sequencing options on a Markov chain. 

\begin{figure}[htbp]
\begin{center}\vspace{0.8cm}
\includegraphics[width=1.0\linewidth]{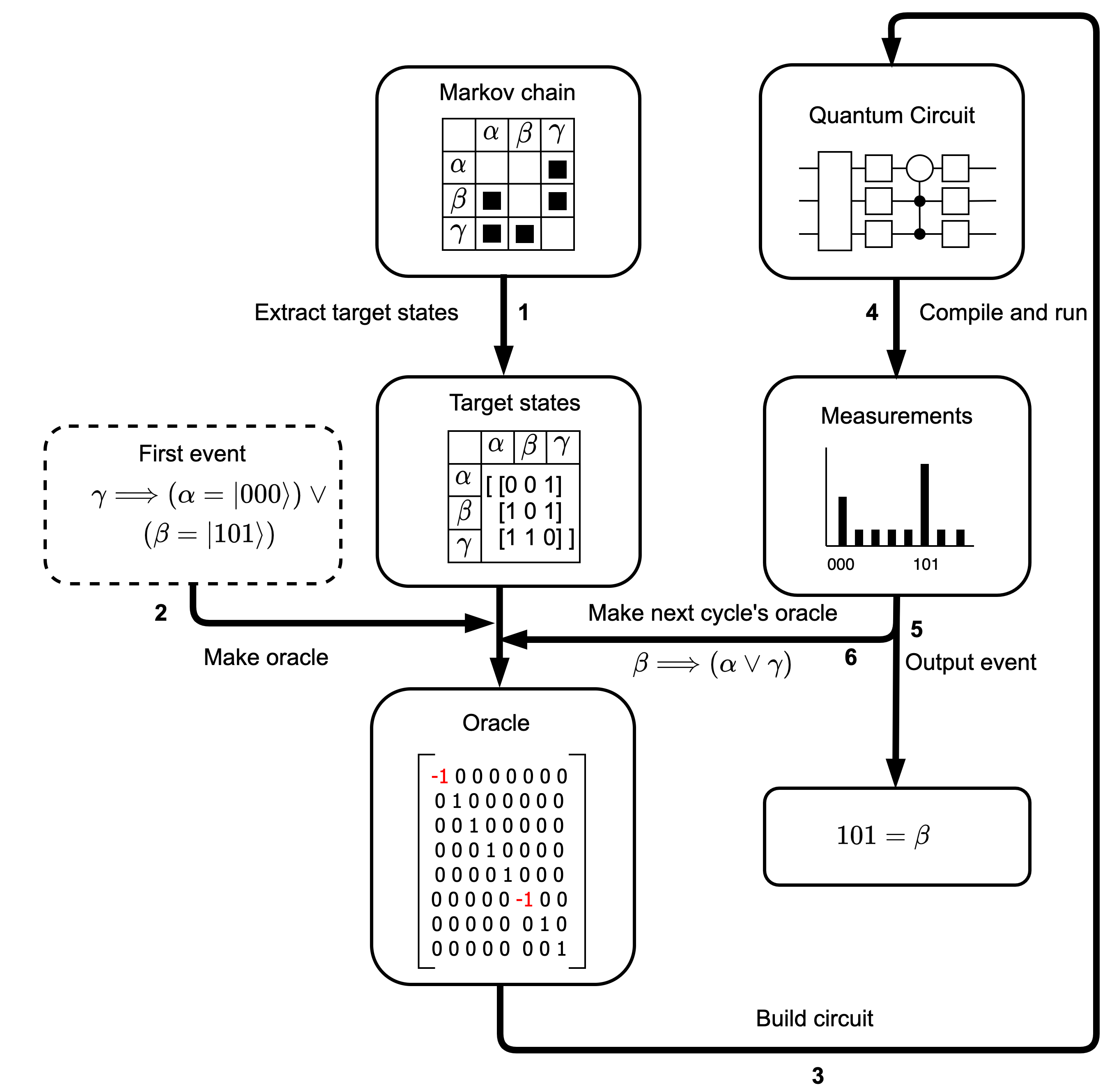}
\caption{The Basak-Miranda algorithm.}
\label{fig:BasMir-algorithm}
\end{center}
\end{figure}

\medskip
The flow diagram of the Basak-Miranda algorithm is shown in Figure \ref{fig:BasMir-algorithm}. Let us first have a bird's-eye view of the algorithm by means of table trace example. Then, we will focus on some of its important details.

\medskip
Given the Markov chain representing the sequencing rules shown in Table \ref{table:rulesMarkov}, the algorithm builds a matrix $\mathcal{T}$ of \textit{target states} (Figure \ref{fig:BasMir-algorithm} (1)), as follows:

\setcounter{MaxMatrixCols}{20}
\begin{equation}
\mathcal{T} = 
\begin{bmatrix}
0 & 1 & 0 & 0 & 0 & 1 & 0 & 0 & 0 & 0 & 0 & 0\\
1 & 0 & 0 & 1 & 1 & 0 & 1 & 0 & 0 & 0 & 0 & 0 \\
0 & 0 & 1 & 0 & 1 & 1 & 0 & 0 & 0 & 0 & 0 & 0 \\
0 & 1 & 0 & 1 & 0 & 0 & 0 & 1 & 0 & 0 & 0 & 0 \\
0 & 1 & 1 & 0 & 0 & 0 & 1 & 0 & 1 & 0 & 0 & 0 \\
1 & 0 & 1 & 0 & 0 & 0 & 0 & 1 & 0 & 1 & 0 & 0 \\
0 & 1 & 0 & 0 & 1 & 0 & 0 & 0 & 1 & 1 & 0 & 0 \\
0 & 0 & 0 & 1 & 0 & 1 & 0 & 0 & 0 & 1 & 0 & 1 \\
0 & 0 & 0 & 0 & 1 & 0 & 1 & 0 & 0 & 1 & 0 & 1 \\
0 & 0 & 0 & 0 & 0 & 1 & 0 & 1 & 1 & 0 & 1 & 0 \\
0 & 0 & 0 & 0 & 0 & 0 & 1 & 1 & 0 & 1 & 0 & 1 \\
0 & 0 & 0 & 0 & 0 & 0 & 0 & 0 & 1 & 0 & 1 & 0
\end{bmatrix}
\end{equation}

\medskip
In fact, Table \ref{table:rulesblocks} comes in handy to clearly visualise what this matrix represents: a black square is a target state, or a digit 1 in the matrix.

\medskip
Next, the algorithm extracts from the matrix the row for the respective rule to be processed. This rule is designated by the last generated note. For instance, if the last note was a D\musSharp{}, then Rule 2 is to be processed. Therefore, the algorithm extracts the second row, which gives the target states for this rule: $R2 = [1 \ 0 \ 0 \ 1 \ 1\  0 \ 1 \ 0 \ 0 \ 0 \ 0 \ 0]$. 

\medskip
Then, the extracted row is converted into a quantum gate, which is referred to as an \textit{oracle} (Figure \ref{fig:BasMir-algorithm} (2)). The notion of oracle will be discussed in more detail later. 

\medskip
Mathematically, quantum states are notated as vectors and quantum gates as matrices. In a nutshell, the oracle is expressed by an identity matrix, whose columns are tensor vectors corresponding to quantum states. Columns corresponding to target states are marked as negative. 

\medskip
The resulting oracle for Rule 2 is shown below. The target states $\ket{0}_4$, $\ket{3}_4$, $\ket{4}_4$, $\ket{6}_4$ (or pitches E, C\musSharp{}, F\musSharp and G\musSharp{}, respectively) are in bold, for clarity\footnote{{\footnotesize The subscript 4 next to the ket indicates the number of qubits required to represent the respective decimal number in binary form.}}.

\setcounter{MaxMatrixCols}{20}
\begin{equation}
\mathcal{O}(R2)= 
\begin{bmatrix}
\textbf{-1} & 0 & 0 & \textbf{0}  & \textbf{0}  & 0 & \textbf{0} & 0 & 0 & 0 & 0 & 0 & 0 & 0 & 0 & 0 \\
\textbf{0}  & 1 & 0 & \textbf{0}  & \textbf{0}  & 0 & \textbf{0} & 0 & 0 & 0 & 0 & 0 & 0 & 0 & 0 & 0 \\
\textbf{0}  & 0 & 1 & \textbf{0}  & \textbf{0}  & 0 & \textbf{0} &  0 & 0 & 0 & 0 & 0 & 0 & 0 & 0 & 0 \\
\textbf{0}  & 0 & 0 & \textbf{-1} & \textbf{0}  & 0 & \textbf{0} & 0 & 0 & 0 & 0 & 0 & 0 & 0 & 0 & 0 \\
\textbf{0}  & 0 & 0 & \textbf{0}  & \textbf{-1} & 0 & \textbf{0} & 0 & 0 & 0 & 0 & 0 & 0 & 0 & 0 & 0 \\
\textbf{0}  & 0 & 0 & \textbf{0}  & \textbf{0}  & 1 & \textbf{0} & 0 & 0 & 0 & 0 & 0 & 0 & 0 & 0 & 0 \\
\textbf{0}  & 0 & 0 & \textbf{0}  & \textbf{0}  & 0 & \textbf{-1} & 0 & 0 & 0 & 0 & 0 & 0 & 0 & 0 & 0 \\
\textbf{0}  & 0 & 0 & \textbf{0}  & \textbf{0}  & 0 & \textbf{0} & 1 & 0 & 0 & 0 & 0 & 0 & 0 & 0 & 0 \\
\textbf{0}  & 0 & 0 & \textbf{0}  & \textbf{0}  & 0 & \textbf{0} & 0 & 1 & 0 & 0 & 0 & 0 & 0 & 0 & 0 \\
\textbf{0}  & 0 & 0 & \textbf{0}  & \textbf{0}  & 0 & \textbf{0} & 0 & 0 & 1 & 0 & 0 & 0 & 0 & 0 & 0 \\
\textbf{0}  & 0 & 0 & \textbf{0}  & \textbf{0}  & 0 & \textbf{0} & 0 & 0 & 0 & 1 & 0 & 0 & 0 & 0 & 0 \\
\textbf{0}  & 0 & 0 & \textbf{0}  & \textbf{0}  & 0 & \textbf{0} & 0 & 0 & 0 & 0 & 1 & 0 & 0 & 0 & 0 \\
\textbf{0}  & 0 & 0 & \textbf{0}  & \textbf{0}  & 0 & \textbf{0} & 0 & 0 & 0 & 0 & 0 & 1 & 0 & 0 & 0 \\
\textbf{0}  & 0 & 0 & \textbf{0}  & \textbf{0}  & 0 & \textbf{0} & 0 & 0 & 0 & 0 & 0 & 0 & 1 & 0 & 0 \\
\textbf{0}  & 0 & 0 & \textbf{0}  & \textbf{0}  & 0 & \textbf{0} & 0 & 0 & 0 & 0 & 0 & 0 & 0 & 1 & 0 \\
\textbf{0}  & 0 & 0 & \textbf{0}  & \textbf{0}  & 0 & \textbf{0} & 0 & 0 & 0 & 0 & 0 & 0 & 0 & 0 & 1
\end{bmatrix}
\label{eq:oracle}
\end{equation}

\medskip
In principle our oracle would not need more than 12 quantum states, or columns, since our Markov chain specifies transitions between 12 pitches. However, we need 4 qubits to encode these 12 states because 3 qubits can encode only up to 8 states. Hence we need a 16$\times$16 matrix to express our oracle.

\medskip
The oracle is the first component of the quantum circuit that will select the next note. The other component is the so-called \textit{amplitude amplification} (or amplitude remixing); again, this will be explained in more detail below. The algorithm then assembles the circuit, compiles it and runs the program (Figure \ref{fig:BasMir-algorithm} (3) (4)). 

\medskip
The result will be a target state corresponding to one of the notes allowed by Rule 2: E, C\musSharp{}, F\musSharp{}, or G\musSharp{}. Then, the algorithm outputs the respective note (Figure \ref{fig:BasMir-algorithm} (5)), which in turn designates the rule for generating the next note. Next, it builds a new oracle (Figure \ref{fig:BasMir-algorithm} (6)), assembles the quantum circuit, and so on. This cycle continues for as long as required to generate a composition.

\subsection{Constructive and Destructive Interference}

\medskip
Interference, together with superposition and entanglement, are important characteristics of quantum computing, which make it different from traditional computing.

\medskip
Interference is explained here in the context an algorithm devised by Lov Grover in 1997 \cite{Grover1997}. Grover's algorithm has become a favoured example to demonstrate the superiority of quantum computing for searching databases. However, this application is a little misleading. To perform a real quantum search in a data base, the data would need to be represented as a superposition of quantum states. Moreover, this representation needs to be created at nontrivial high speeds and be readily available for processing though some sort of quantum version of random access memory (RAM). Although the notion of quantum random access memory (QRAM) has been proposed \cite{Giovannetti2008}, they are not practically available at the time of writing.

\medskip
Grover's algorithm is better thought of as an algorithm that can tell us whether something is in a dataset or not. Metaphorically, think of the puzzle books \textit{Where's Wally?} (or \textit{Where's Waldo?} in the USA) \cite{Handford2007}. The books consist of illustrations depicting hundreds of characters doing a variety of amusing things in busy scenarios. Readers are challenged to find a specific character, named Wally, hidden in the illustrations. Assuming that it would be possible to represent those illustrations as quantum states in superposition, then Grover's algorithm would be able to tell us rather quickly whether or not Wally is in there. 

\subsubsection{Oracle}

\medskip
Mathematically, the \textit{Where's Wally} puzzle can be formalised as a problem of finding the unique input(s) to a given function that produces a required result. 

\medskip
Let us formalise a simple musical example: assuming that the following binary string $s = 10100101$ represents the musical note A5, the function $f(x)$ returns 1 when $x = s$, and 0 otherwise:

\begin{align}
f(x) =
  \begin{cases*}
    1 \quad& if  $ x = 10100101 \equiv \text{{\footnotesize A5}} $ \\
    0 & if \text{  {\footnotesize otherwise}} \\
  \end{cases*}
\end{align}
 
\medskip
For quantum computing, the function above is best represented as:

\begin{align}
f(\ket{\Psi}) =
  \begin{cases*}
    \ket{1} \quad& if  $ \ket{\Psi} = \ket{10100101} \equiv \ket{\text{{\footnotesize A5}}} $ \\
    \ket{0} & if \text{  {\footnotesize otherwise}} \\
  \end{cases*}
\end{align}

\medskip
In order to use a quantum computer to find out if, say, a musical composition contains the note A5, all of its notes would need to be provided in superposition to the function. 

\medskip
The standard method for conducting a search for the note A5 is to check every note one-by-one. However, a quantum computer running Grover's algorithm would be able to return a result with considerably lower number of queries.

\medskip
Suppose that the composition in question has 1,000 notes. The best case scenario would be when the first note that the system checks is the one it is looking for. Conversely, the worse case would be when the note it is looking for is the last one checked. In this case, the system would have made 1,000 queries to give us an answer. On average, we could say that the system would have to make ${N}/{2}$ queries; that is, with $N = 1000$, then ${1000}/{2} = 500$ queries.  Conversely, Grover's algorithm would require $\sqrt{N}$ queries; i.e., $\sqrt{1000} \approx 31$ queries. This is significantly faster.

\medskip
In practice, for a quantum computer we need to turn $f(\ket{\Psi})$ into a quantum gate $\mathrm{\mathbf{U}}_f$ describing a unitary linear transformation. 

\begin{align}
\mathrm{\mathbf{U}}_{f{(\ket{\Psi})}} \ket{x}_7 =
  \begin{cases*}
    -\ket{y}_7 \quad& if  $ \ket{x}_7 = \ket{y}_7 \ \ (\text{{\footnotesize which is to say, }} f(\ket{y}_7) = 1 \  \text{{\footnotesize for A5}})$ \\
    \ket{x}_7 & if \text{  {\footnotesize otherwise}} \\
  \end{cases*}
\end{align}

\medskip
The gate $\mathrm{\mathbf{U}}_f$ is what we referred to earlier as an oracle. Note that the oracle marks the target state by flipping the sign of the input that satisfies the condition. The rest is left unchanged. The reason for doing this will become clearer below. Effectively, the oracle $\mathcal{O}(R2)$ in Eq.\ref{eq:oracle} yields a $\mathrm{\mathbf{U}}_f$ with 4 target states.

\subsubsection{Amplitude Remixing}

\medskip
As a didactic example to illustrate how amplitude amplification works, let us consider a two-qubit quantum system $\ket{\phi}$, which gives 4 states on the standard basis, corresponding to 4 possible outcomes:

\begin{equation}
	\ket{00}= 
	\begin{bmatrix} 1 \\ 0 \\ 0 \\ 0
	\end{bmatrix}, \ \ 
	\ket{01}= 
	\begin{bmatrix} 0 \\ 1 \\ 0 \\ 0
	\end{bmatrix}, \ \	
	\ket{10}= 
	\begin{bmatrix} 0 \\ 0 \\ 1 \\ 0
	\end{bmatrix}, \ \
	\ket{11}= 
	\begin{bmatrix} 0 \\ 0 \\ 0 \\ 1
	\end{bmatrix}
\end{equation}

\medskip
If we put the qubits in balanced superposition, then the amplitudes of all possible states will be equally distributed; i.e., ${1}/{\sqrt{2^2}}$ each, which gives $\left|{1}/{\sqrt{2^2}}\right|^2 = 0.25$. In other words, we would end up with an equal probability of 25\% of obtaining any of the outcomes:

\begin{align}
	\mathrm{\mathbf{H}}^{\otimes 2} \ket{\phi} \Rightarrow  \ket{{\phi}_1} = \left[ \frac{1}{\sqrt{2^2}}\ket{00} + \frac{1}{\sqrt{2^2}}\ket{01} + \frac{1}{\sqrt{2^2}}\ket{10} + \frac{1}{\sqrt{2^2}}\ket{11} \right]
\label{eq:bal_sup}
\end{align}

\medskip

\begin{figure}[htbp]
\begin{center}\vspace{0.3cm}
\includegraphics[width=0.58\linewidth]{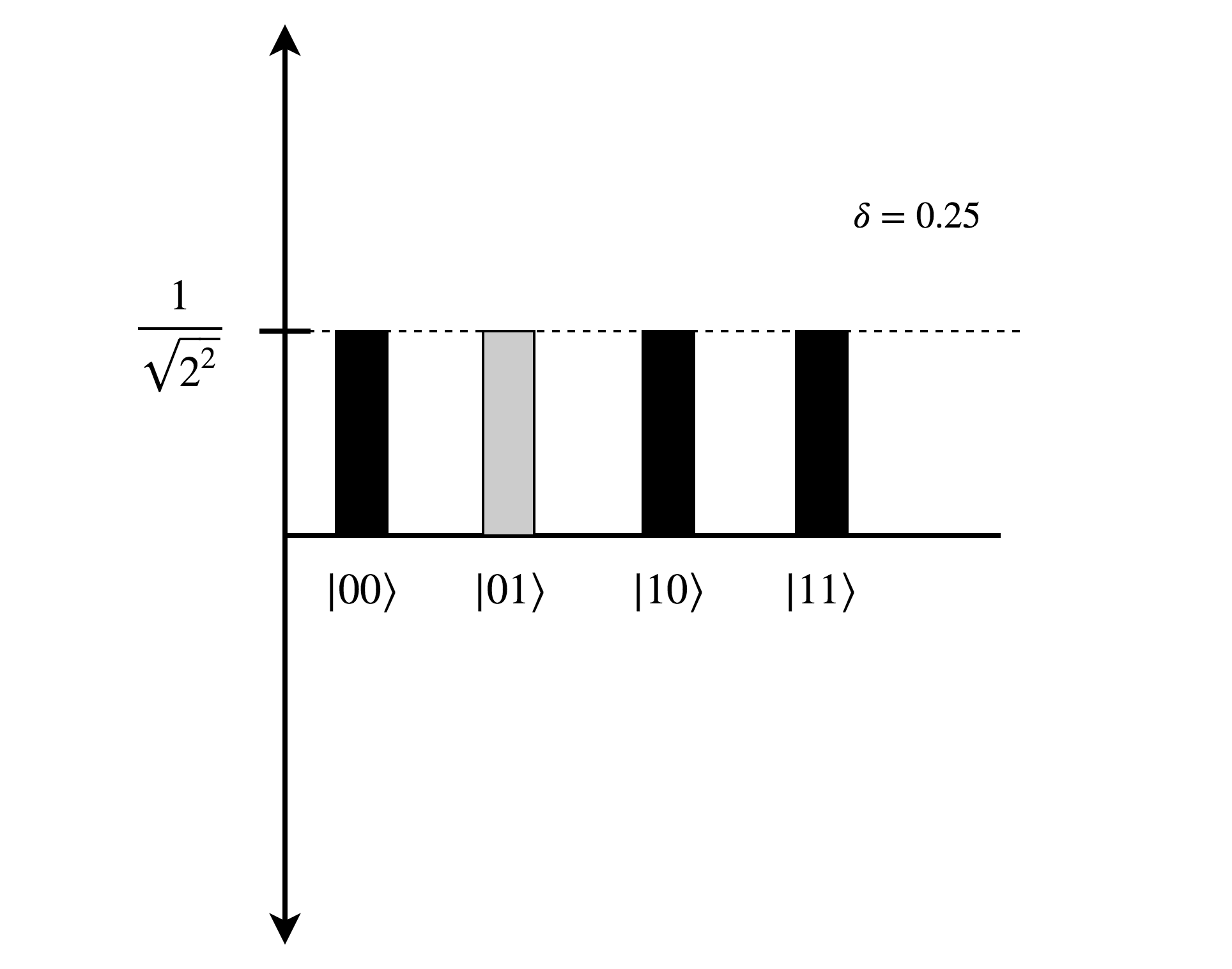}
\caption{Two-qubit system in balanced superposition.}
\label{fig:amp_graph1}
\end{center}
\end{figure}

The graph in Figure \ref{fig:amp_graph1} depicts a visual representation of this balanced superposition. The variable $\delta$ is the squared average amplitude: $\delta = \left|{1}/{\sqrt{2^2}}\right|^2 = 0.25$.

\medskip
 In more detail:

\begin{align}
	\mathrm{\mathbf{H}}^{\otimes 2} = 
	\mathrm{\mathbf{H}} \otimes \mathrm{\mathbf{H}} = 
	\frac{1}{\sqrt{2}}
	\begin{bmatrix} 
	1 & 1 \\
	1 & - 1
	\end{bmatrix} \bigotimes
	\frac{1}{\sqrt{2}}
	\begin{bmatrix} 
	1 & 1 \\
	1 & - 1
	\end{bmatrix} =
	\frac{1}{\sqrt{2^2}}
	\begin{bmatrix} 
	1 & 1 & 1 & 1\\
	1 & - 1 & 1 & -1 \\
	1 & 1 & -1 & -1 \\
	1 & -1 & -1 & 1
	\end{bmatrix}	
\end{align}

\medskip

Therefore:

\begin{equation}
\label{eqn:equal_sup}
	\mathrm{\mathbf{H}}^{\otimes 2} \ket{00} = 
	\frac{1}{\sqrt{2^2}}
	\begin{bmatrix} 
		1 & 1 & 1 & 1\\
		1 & - 1 & 1 & -1 \\
		1 & 1 & -1 & -1 \\
		1 & -1 & -1 & 1
	\end{bmatrix}
	\begin{bmatrix} 1 \\ 0 \\ 0 \\ 0 \end{bmatrix} =
	\frac{1}{\sqrt{2^2}}	
	\begin{bmatrix} 1 \\ 1 \\ 1 \\ 1 \end{bmatrix}
\end{equation}

\medskip
Let us suppose that the target state that we are looking for is $\ket{01}$. The oracle $\mathcal{O}$ to mark this target would look like this:

\begin{align}
	\mathcal{O} = 
	\begin{bmatrix} 
		1 & \mathbf{0} & 0 & 0\\
		0 & \mathbf{- 1} & 0 & 0 \\
		0 & \mathbf{0} & 1 & 0 \\
		0 & \mathbf{0} & 0 & 1
	\end{bmatrix}
\end{align}

If the apply the oracle $\mathcal{O}$ to mark the target on $\ket{{\phi}_1}$ (Eq.\ref{eq:bal_sup}), then the target's signal will be flipped. That is, its amplitude will be reversed:

\begin{equation}
\label{eqn:aply_oracle}
\begin{aligned}
	\mathcal{O}\ket{{\phi}_1} \Rightarrow & \ket{{\phi}_2} = \left[ \frac{1}{\sqrt{2^2}}\ket{00} \left(- \frac{1}{\sqrt{2^2}}\ket{01}\right) + \frac{1}{\sqrt{2^2}}\ket{10} + \frac{1}{\sqrt{2^2}}\ket{11} \right] \\ \\
	& \ket{{\phi}_2} = \frac{1}{\sqrt{2^2}} \left[ \ket{00} - \ket{01} + \ket{10} + \ket{11} \right]
\end{aligned}	
\end{equation}

\begin{figure}[htbp]
\begin{center}\vspace{0.3cm}
\includegraphics[width=0.5\linewidth]{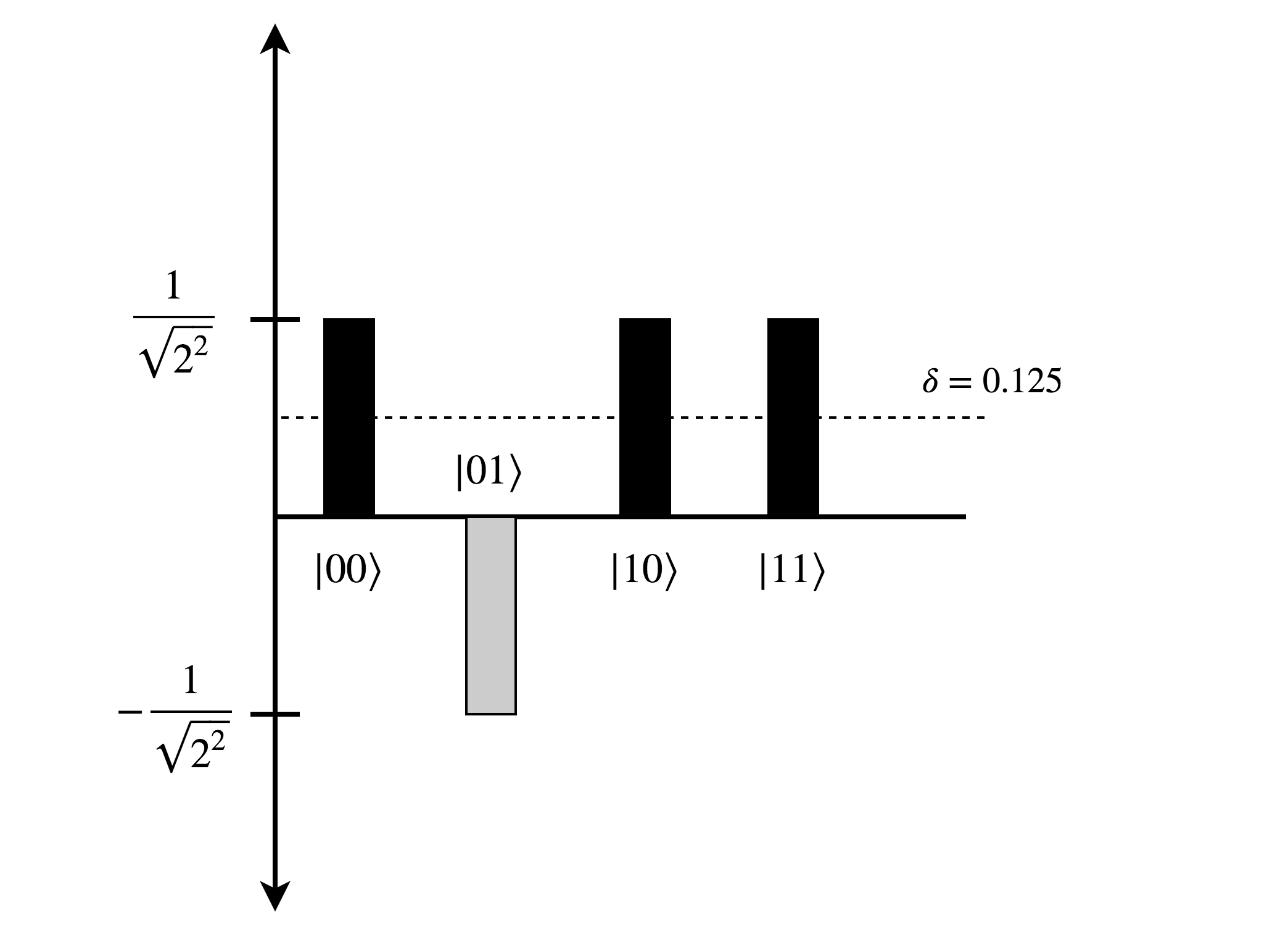}
\caption{The oracle inverts the amplitude of the target state.}
\label{fig:amp_graph2}
\end{center}
\end{figure}

\medskip
The graph in Figure \ref{fig:amp_graph2} illustrates the effect of reversing the amplitude. The squared average has halved: $\delta = \left|{1}/{\sqrt{2.82^2}}\right|^2 \approx 0.125$.

\medskip
In more detail:

\begin{align}
	\mathcal{O}\ket{{\phi}_1} \Rightarrow & \ket{{\phi}_2} = 
	\begin{bmatrix} 
		1 & \mathbf{0} & 0 & 0\\
		0 & \mathbf{- 1} & 0 & 0 \\
		0 & \mathbf{0} & 1 & 0 \\
		0 & \mathbf{0} & 0 & 1
	\end{bmatrix}
	\frac{1}{\sqrt{2^2}}	
	\begin{bmatrix} 1 \\ 1 \\ 1 \\ 1 \end{bmatrix}	 =
	\frac{1}{\sqrt{2^2}} \begin{bmatrix} 1 \\ -1 \\ 1 \\ 1 \end{bmatrix}	
	\ \ \text{or}	 \ \
	\begin{bmatrix} {1}/{\sqrt{2^2}} \\ -{1}/{\sqrt{2^2}} \\ {1}/{\sqrt{2^2}} \\ {1}/{\sqrt{2^2}} \end{bmatrix}
\end{align}

\medskip
Obviously, reversing the amplitude of the target still does not change the equal chance of obtaining any of the 4 outcomes. It is at this stage that amplitude amplification is put into action. What it does is to unbalance the amplitudes of the quantum system: it will augment the amplitude of the marked target state and decrease the amplitudes of all others. Thus, when we measure the system, now we would almost certainly obtain the target. Hence, we propose to refer to this procedure as \textit{amplitude remixing} rather than amplitude amplification; it is a more intuitive term for a musician.

\medskip
An amplitude remixer is defined as a unitary linear transformation, or gate $\mathrm{\mathbf{U}}_\varphi$, as follows: 

\begin{equation}
\label{eqn:amp_amp}
\mathrm{\mathbf{U}}_\varphi = \mathrm{\mathbf{H}}^{\otimes n} \ \mathrm{\mathbf{S}} \ \mathrm{\mathbf{H}}^{\otimes n}
\end{equation}

\medskip
The operator $\mathrm{\mathbf{S}}$ acts as a conditional shift matrix operator of the form:

\begin{align}
	\mathrm{\mathbf{S}}= 
	\begin{bmatrix} 
		1 & 0 & 0 & 0\\
		0 & e^{i\pi} & 0 & 0 \\
		0 & 0 & e^{i\pi} & 0 \\
		0 & 0 & 0 & e^{i\pi} 
	\end{bmatrix} 
	=
	\begin{bmatrix} 
		1 & 0 & 0 & 0\\
		0 & -1 & 0 & 0 \\
		0 & 0 & -1 & 0 \\
		0 & 0 & 0 & -1
	\end{bmatrix}
\end{align}

\medskip
Thus, the application of $\mathrm{\mathbf{U}}_\varphi$ to $\ket{{\phi}_2}$ looks like this:

\begin{equation}
\label{eqn:aply_ampamp}
\begin{aligned}
	 \mathrm{\mathbf{U}}_\varphi \ket{{\phi}_2} \Rightarrow & \mathrm{\mathbf{H}}^{\otimes n} \ket{{\phi}_2} \rightarrow \ket{{\phi}_3} = 
	 \frac{1}{\sqrt{2^2}}
	 \begin{bmatrix} 
		1 & 1 & 1 & 1\\
		1 & - 1 & 1 & -1 \\
		1 & 1 & -1 & -1 \\
		1 & -1 & -1 & 1
	\end{bmatrix}
	 \frac{1}{\sqrt{2^2}} \begin{bmatrix} 1 \\ -1 \\ 1 \\ 1 \end{bmatrix}	=
	 \frac{1}{\sqrt{4^2}} \begin{bmatrix} 2 \\ 2 \\ -2 \\ 2 \end{bmatrix}  = 
	 \frac{1}{\sqrt{2^2}} \begin{bmatrix} 1 \\ 1 \\ -1 \\ 1 \end{bmatrix}  \\ \\
	& \mathrm{\mathbf{S}} \ket{{\phi}_3} \rightarrow \ket{{\phi}_4} =
	\begin{bmatrix} 
		1 & 0 & 0 & 0\\
		0 & -1 & 0 & 0 \\
		0 & 0 & -1 & 0 \\
		0 & 0 & 0 & -1
	\end{bmatrix}
	\frac{1}{\sqrt{2^2}} \begin{bmatrix} 1 \\ 1 \\ -1 \\ 1 \end{bmatrix} =
	\frac{1}{\sqrt{2^2}} \begin{bmatrix} 1 \\ -1 \\ 1 \\ -1 \end{bmatrix} 	\\ \\
	& \mathrm{\mathbf{H}}^{\otimes n} \ket{{\phi}_4} \rightarrow \ket{{\phi}_5} =
	\frac{1}{\sqrt{2^2}}
	 \begin{bmatrix} 
		1 & 1 & 1 & 1\\
		1 & - 1 & 1 & -1 \\
		1 & 1 & -1 & -1 \\
		1 & -1 & -1 & 1
	\end{bmatrix}
	\frac{1}{\sqrt{2^2}} \begin{bmatrix} 1 \\ -1 \\ 1 \\ -1 \end{bmatrix} =
	\frac{1}{\sqrt{4^2}} \begin{bmatrix} 0 \\ 4 \\ 0 \\ 0 \end{bmatrix} =
	1 \begin{bmatrix} 0 \\ 1 \\ 0 \\ 0 \end{bmatrix}
\end{aligned}	
\end{equation}

\begin{figure}[htbp]
\begin{center}\vspace{0.3cm}
\includegraphics[width=0.5\linewidth]{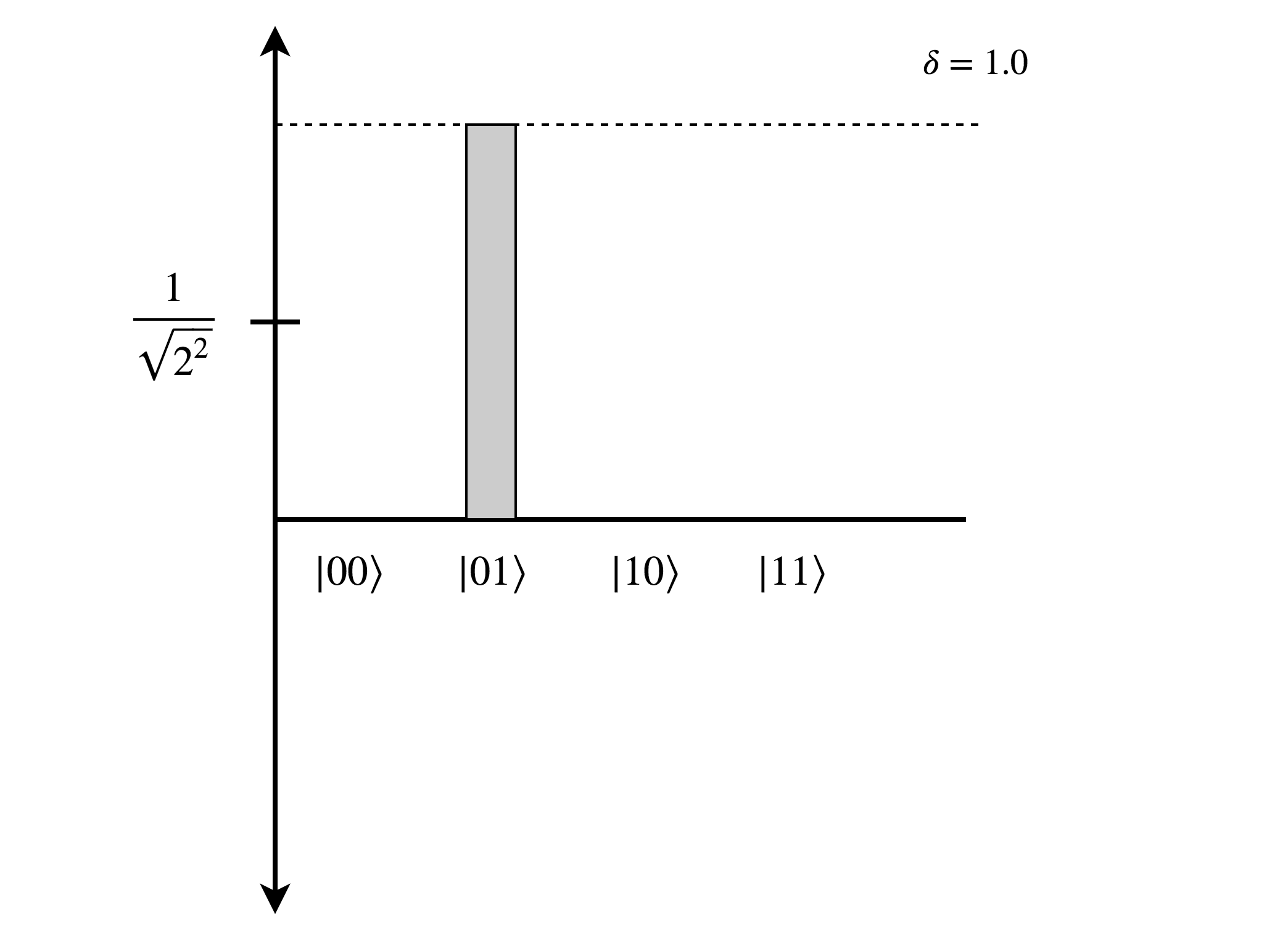}
\caption{There is now a 100\% chance for outcome $\ket{01}$.}
\label{fig:amp_graph3}
\end{center}
\end{figure}

\medskip
So, $\ket{{\phi}_5} = 1.0\ket{01}$, which gives $\left|1\right|^2 = 1$, or in other words, there is now a 100\% chance for outcome $\ket{01}$. The graph in Figure \ref{fig:amp_graph3} shows what happened with the amplitudes of our two-qubit system. The quantum computing literature often refers to this effect as an \textit{inversion about the mean}.

\medskip
To summarize, the transformation $\mathrm{\mathbf{U}}_{f{(\ket{\Psi})}}\mathrm{\mathbf{U}}_\varphi\ket{\phi}$ transforms $\ket{\phi}$ towards a target state $\ket{\Psi}$. This worked well for the above didactic two-qubit system. But for systems with higher number of qubits it is necessary to apply the transformation a number of times. The necessary number of iterations is calculated as: $i = \lfloor \frac{\pi\sqrt{\frac{2^n}{T}}}{4} \rfloor \approx \lfloor 0.7854 \sqrt{\frac{2^n}{T}} \rfloor$, where $n$ is the number of qubits and $T$ is the number of targets. Thus, for 2 qubits and 1 target, $i = \lfloor 0.7854\sqrt{\frac{2^2}{1}} \rfloor =  \lfloor 0.7854\sqrt{4} \rfloor = 1$. And also, $i = 1$ for the Basak-Miranda algorithm example for Rule 2, with 4 target states, as discussed earlier: $i = \lfloor 0.7854\sqrt{\frac{2^4}{4}} \rfloor = 1$. (We use the floor of $i$, generally because the floor requires a shallower circuit. In case of $i < 1$ then $i = 1$.) Should the example above have required more iterations, the \say{Oracle} and \say{Quantum Circuit} stages of the block diagram in Figure \ref{fig:BasMir-algorithm} would have to be repeated accordingly. 

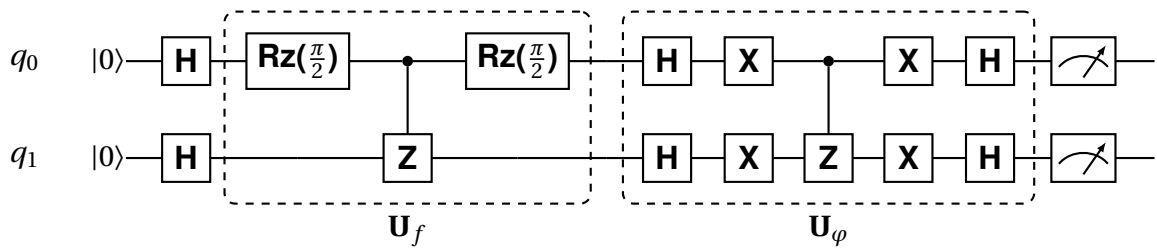
\begin{figure}[htbp]
\begin{center}\vspace{0.3cm}
    \begin{tikzpicture}
        \node[scale=1.0] 
        {
            \begin{quantikz}
                \lstick{$q_0$} &  \ket{0} &  \gate{\textbf{H}} & \gate{\textbf{Rz($\frac{\pi}{2}$)}}\gategroup[wires=2, steps=3, style={dashed, rounded corners}, label style={label position=below, anchor=north, yshift=-0.2cm}] {$\mathrm{\mathbf{U}}_f$}  & \ctrl{1}  & \gate{\textbf{Rz($\frac{\pi}{2}$)}} & \qw & \gate{\textbf{H}}\gategroup[wires=2, steps=5, style={dashed, rounded corners, label position=below}, label style={label position=below, anchor=north, yshift=-0.2cm}] {$\mathrm{\mathbf{U}}_\varphi$} & \gate{\textbf{X}} & \ctrl{1} &\gate{\textbf{X}} & \gate{\textbf{H}} & \meter{} & \qw \\
                \lstick{$q_1$} &  \ket{0} &  \gate{\textbf{H}}  & \qw & \gate{\textbf{Z}} & \qw	& \qw & \gate{\textbf{H}} & \gate{\textbf{X}} & \gate{\textbf{\textbf{Z}}} & \gate{\textbf{X}} & \gate{\textbf{H}} & \meter{} & \qw
            \end{quantikz}
        };
    \end{tikzpicture}
\end{center}
\caption{Quantum circuit implementation of the $\mathrm{\mathbf{U}}_{f}\mathrm{\mathbf{U}\varphi}\ket{\phi}$ interference transformation ($\mathrm{\mathbf{Z}}$ is $\mathrm{\mathbf{Rz(\pi)}}$). }
\label{fig:amp_remix}
\end{figure}

\medskip
A quantum circuit implementing the transformation $\mathrm{\mathbf{U}}_{f}\mathrm{\mathbf{U}\varphi}\ket{\phi}$ discussed above is shown in Figure \ref{fig:amp_remix}. The same architecture applies for the Basak-Miranda algorithm depicted in Figure \ref{fig:BasMir-algorithm}: one just needs to stack two additional qubits to the circuit. And, of course, the oracle $\mathrm{\mathbf{U}}_f$ is replaced each time a different rule is evaluated.

%
\subsection{A Simple Example}
%

\begin{figure}[htbp]
\begin{center}\vspace{0.3cm}
\includegraphics[width=0.5\linewidth]{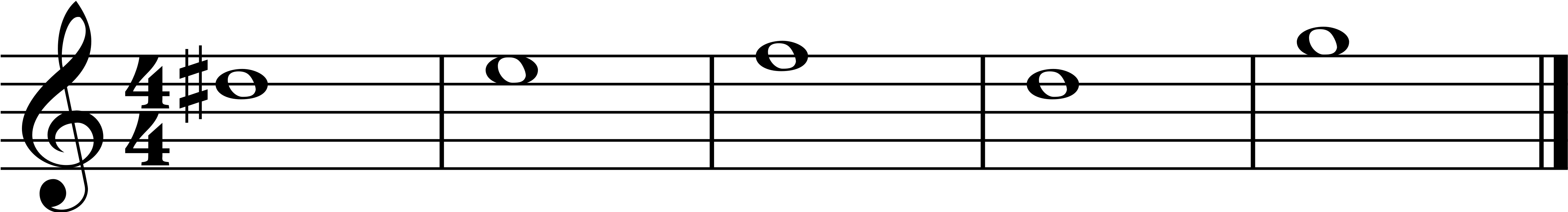}
\caption{Musical output yielded by the the Basak-Miranda algorithm.}
\label{fig:bas-mir-music}
\end{center}
\end{figure}

\medskip
A simple illustrative example of an outcome from the Basak-Miranda algorithm acting on Table \ref{table:rulesMarkov} is shown in Figure \ref{fig:bas-mir-music}. In this example, rhythm is not considered; hence all notes are of the same duration. 

\begin{figure}[htbp]
\begin{center}\vspace{0.3cm}
\includegraphics[width=0.86\linewidth]{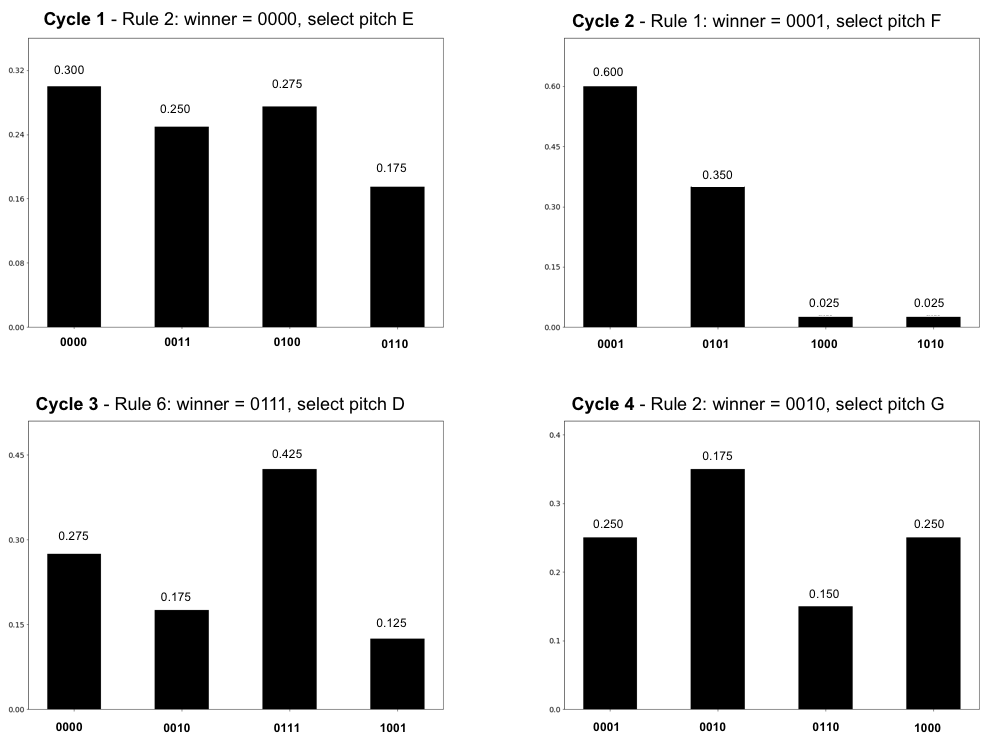}
\caption{Histograms from running the Basak-Miranda algorithm for 4 cycles. The vertical coordinate is the percentage of times an item listed on the horizontal coordinate occurred.}
\label{fig:bas-mir-example}
\end{center}
\end{figure}

\medskip
Consider the histograms in Figure \ref{fig:bas-mir-example}, showing four cycles of the algorithm.  As with the 3-D musical random walk example, 40 shots for each cycle were sufficient here to obtain plausible results\footnote{For this example we used IBM Quantum's QASM simulator.}.

\medskip
The initial pitch is D$\musSharp{}$. Therefore, the system starts by processing Rule 2, which stipulates that one of these 4 pitches allowed allowed next: E, C$\musSharp{}$, F$\musSharp{}$ or G$\musSharp{}$. That is, the target states are $\ket{0}_4$, $\ket{3}_4$, $\ket{4}_4$ and $\ket{6}_4$. In this instance, the algorithm picked pitch E, because this is the option that would have been measured most frequently; i.e., the winner is $\ket{0}_4$, represented as $0000$ on the histogram (Figure \ref{fig:bas-mir-example}, Cycle 1). 

\medskip
Now, the current pitch is E. So, the rule to be processed next is Rule 1. According to this rule, only pitches F or D$\musSharp{}$ can follow E. The target states are $\ket{1}_4$ and $\ket{5}_4$. And in this case, the system picked pitch F, because the winner state would have been $\ket{1}_4$ ($0001$ on the histogram in Figure \ref{fig:bas-mir-example}, Cycle 2).

\medskip
Next,  the current pitch is F. And so, the system processes Rule 6, which specifies 4 possible pitches: E, G, D or C. That is, the target states are $\ket{0}_4$, $\ket{2}_4$, $\ket{7}_4$ and $\ket{9}_4$. The winner state is $\ket{7}_4$, or $0111$ on the histogram (Figure \ref{fig:bas-mir-example}, Cycle 3). And so on.

%
%

\section{Concluding Discussion}

\medskip
Quantum computing is a nascent technology. But one that is advancing rapidly. 

\medskip
There is a long history of research into using computers for making music. Nowadays, computers are absolutely essential for the music economy. Therefore, it is very likely that quantum computers will impact in the music industry in time to come. A new area of research and development is emerging: \textit{Quantum Computer Music}. This chapter laid the foundations of this new exciting field. 

\medskip
In many ways, in terms of development stage, it is fair compare state-of-the-art quantum computers with the computer mainframes of the 1950s. It is hard to imagine how computers will be like in 2090, in the same way that it must have been hard for our forefathers of the 1950s to imagine how computers look like today. But in contrast to 70 years ago, today’s musicians are generally conversant with computing technology. Hence we believe that the time is ripe to start experimenting with quantum computers in music.  

\medskip
A sensible entryway in this fascinating field is to revisit tried-and-tested computer music methods and repurpose them for quantum computing. As one becomes increasingly more acquainted with core concepts and techniques, new quantum-specific algorithms for music will emerge. A example of this  is the Basak-Miranda algorithm introduced above, probably the first ever bespoke quantum algorithm for generating music.

\medskip
As compared to an ordinary laptop computer, it must be acknowledged that there is absolutely no advantage of running the systems introduced in this chapter on a quantum computer. The problems are somewhat trivial. And quantum computing hardware is not yet available to tackle complex music problems. Indeed, all the examples above were run on software simulations of quantum hardware. At this stage, we are not advocating any quantum advantage for musical applications. What we advocate, however, is that the music technology community should be quantum-ready for when quantum computing hardware becomes more sophisticated, widely available, and possibly advantageous for creativity and business. Nevertheless, we propose   that quantum computing is bringing two benefits to music: (a) a new paradigm for creativity and (b) algorithm speed-up.

\medskip
It is a fact that computing has always been influencing how musicians create music \cite{Roads2015}. For instance, there are certain styles of music that would have never been created if programming languages did not exist. Indeed, there are music genres nowadays where the audience watches musicians programming live on stage (as if they were playing musical instruments), and night clubs where the DJs write code live instead of spinning records \cite{Calore2019}. We anticipate that new ways of thinking boosted by quantum programming, and the modelling and simulations afforded by quantum computing, will constitute a new paradigm for creativity, which would not have been emerged otherwise. An example of this is the composition \textit{Zeno} by the first author \cite{Miranda2020a} \cite{Miranda2020b}.

\medskip
As for algorithm speed-up, it is too early to corroborate any hypothesis one could possibly make at this stage. It has been argued that quantum walk on real quantum hardware would be faster than classical random walk to navigate vast mathematical spaces \cite{Kendon2006}.  Quantum walk is an area of much interest for computer music. 

\medskip
Moreover, we hypothesise that the Basak-Miranda algorithm would work faster on quantum computers than on classical ones for considerable large chains and higher number of target states.  As discussed earlier, the interference technique is at the core of Grover’s algorithm. To check for an element in an unstructured set of $N$ elements, a brute-force classic algorithm would scan all elements in the set until it finds the one that is sought after. In the worst-case scenario, the element in question could have been the last one to be checked, which means that the algorithm would have made $N$ queries to find it. Grover’s algorithm would be able find a solution with $\sqrt{N}$ queries. Thus, the algorithm provides a quadratic speedup. Theoretically, this benchmarking is applicable to assess the Basak-Miranda algorithm. But this must be verified in practice when suitable hardware becomes available, as there might be other factors to be considered; e.g., the extent to which ancillary classical processing is required alongside quantum processing to implement a practical system.

%
%

\section*{Appendix: An Introduction to Gate-Based Quantum Computing}

\medskip

This brief introduction to gate-based quantum computing builds upon an extract from \cite{Miranda2021}. It introduces only the basics deemed necessary to follow the discussions in this chapter. The reader is referred to \cite{Bernhardt2019} \cite{Grumbling2019} \cite{Rieffel2011} \cite{Sutor2019} for more detailed explanations.

\medskip
Classical computers manipulate information represented in terms of binary digits, each of which can value 1 or 0. They work with microprocessors made up of billions of tiny switches that are activated by electric signals. Values 1 and 0 reflect the on and off states of the switches. 

\medskip
In contrast, a quantum computer deals with information in terms of quantum bits, or \textit{qubits}. Qubits operate at the subatomic level. Therefore, they are subject to the laws of quantum physics. 

\medskip
At the subatomic level, a quantum object does not exist in a determined state. Its state is unknown until one observes it. Before it is observed, a quantum object is said to behave like a wave. But when it is observed it becomes a particle. This phenomenon is referred to as the wave-particle duality.

\medskip
Quantum systems are described in terms of wave functions. A wave function represents what the particle would be like when a quantum object is observed. It expresses the state of a quantum system as the sum of the possible states that it may fall into when it is observed. Each possible component of a wave function, which is also a wave, is scaled by a coefficient reflecting its relative weight. That is, some states might be more likely than others. Metaphorically, think of a quantum system as the audio spectrum of a musical note, where the different amplitudes of its various wave-components give its unique timbre. As with sound waves, quantum wave-components interfere with one another, constructively and destructively. In quantum physics, the interfering waves are said to be coherent. As we will see later, the act of observing waves decoheres them. Again metaphorically, it is as if when listening to a musical note one would perceive only a single spectral component; probably the one with the highest energy, but not necessarily so.

\medskip
Qubits are special because of the wave-particle duality. Qubits can be in an indeterminate state, represented by a wave function, until they are read out. This is known as superposition. A good part of the art of programming a quantum computer involves manipulating qubits to perform operations while they are in such indeterminate state. This makes quantum computing fundamentally different from digital computing. 

\medskip
Qubits can be implemented in a number of ways. Depending on the technology used to make a quantum processor, its qubits  need to be isolated from the environment in order to remain coherent to perform computations. Quantum coherence is a \textit{sine qua non} for the operation of quantum processors. It is what maintains qubits in superposition and enables the kinds of interference discussed in this chapter.  Yet, coherence is doomed to fall apart - i.e., to decohere - before any nontrivial circuit has a chance to run to its end. The promising advantages of quantum computers are crippled by decoherence, which leads to fatal processing errors. Metaphorically, think of keeping qubits coherent as something like balancing a bunch of coins upright on an unstable surface, where any movement, even the tiniest vibration, would cause them to fall to head or tail. In short, any interaction with the environment causes qubits to decohere. But it is extremely hard, if not impossible, to shield a quantum chip from the environment. 

\begin{figure}[htbp]
\begin{center}\vspace{0.3cm}
\includegraphics[width=0.3\linewidth]{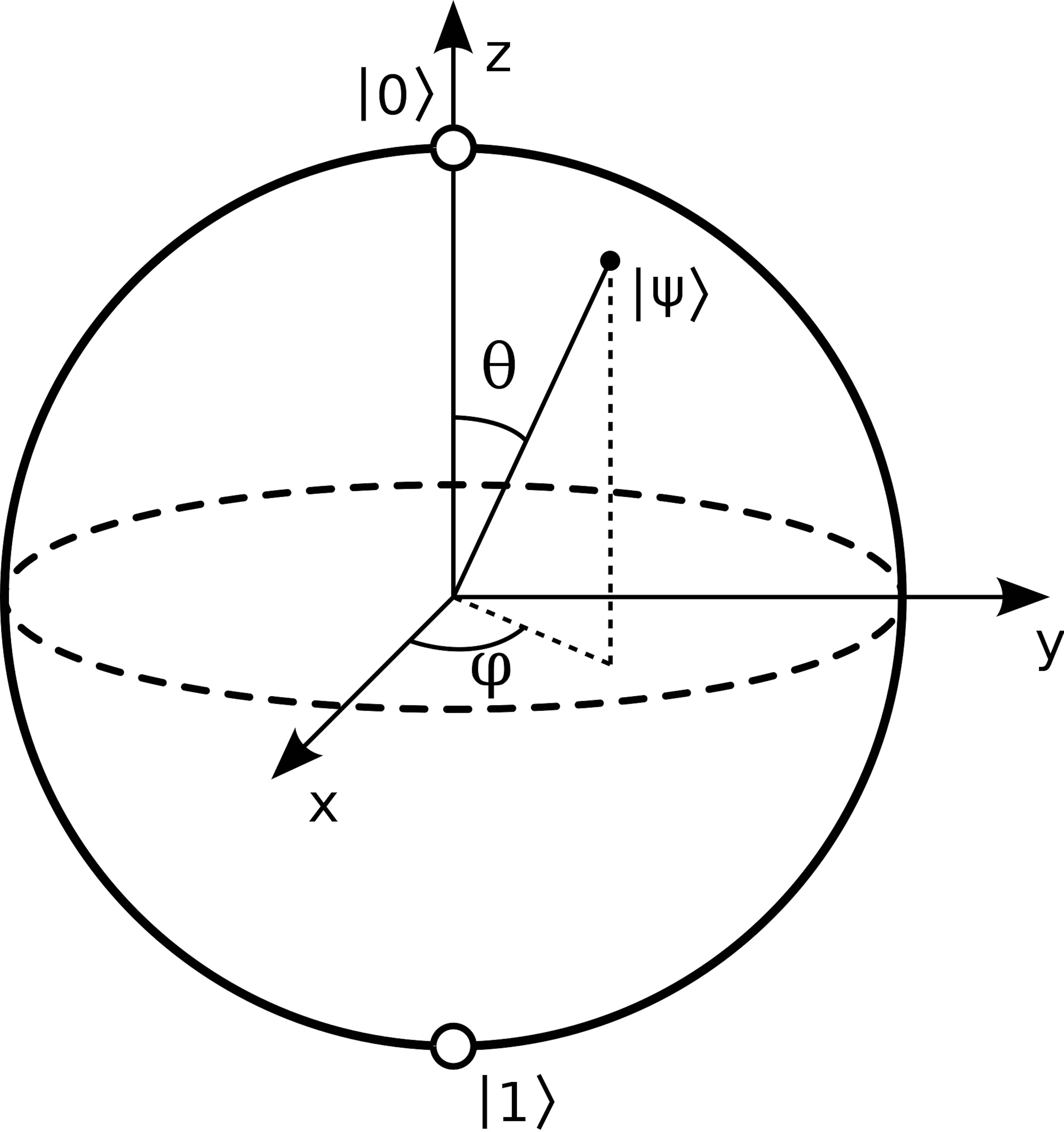}
\caption{Bloch sphere. \\ (Source: Smite-Meister, https://commons.wikimedia.org/w/index.php?curid=5829358)}
\label{fig:Bloch_sphere}
\end{center}
\end{figure}

\medskip
In order to picture a qubit, imagine a transparent sphere with opposite poles. From its centre, a vector whose length is equal to the radius of the sphere can point to anywhere on the surface. In quantum mechanics this sphere is called Bloch sphere and the vector is referred to as a state vector. The opposite poles of the sphere are denoted by $\ket{0}$ and $\ket{1}$, which is the notation used to represent quantum states (Figure \ref{fig:Bloch_sphere}).

\medskip
A qubit’s state vector can point at anywhere on the Bloch sphere’s surface. Mathematically, it is described in terms of polar coordinates using two angles, $\theta$ and $\varphi$. The angle $\theta$ is the angle between the state vector and the z-axis (latitude) and the angle $\varphi$ describes vector’s position in relation to the x-axis (longitude).

\medskip
It is popularly said that a qubit can value 0 and 1 at the same time, but this is not entirely accurate.  When a qubit is in superposition of $\ket{0}$ and $\ket{1}$, the state vector could be pointing anywhere between the two. However, we cannot really know where exactly a state vector is pointing to until we read the qubit. In quantum computing terminology, the act of reading a qubit is referred to as 'observing', or ‘measuring’ it. Measuring the qubit will make the vector point to one of the poles and return either 0 or 1 as a result.

\medskip
The state vector of a qubit in superposition state is described as a linear combination of two vectors,$\ket{0}$ and $\ket{1}$, as follows: 

\begin{equation}
\ket{\Psi} = \alpha \ket{0} + \ket{1},  \ \ \text{where} \ \  \left| \alpha \right|^2=0.5 \ \ \text{and} \ \ \left| \beta \right|^2=0.5
\end{equation}

\medskip
The state vector $\ket{\Psi}$ is a superposition of vectors $\ket{0}$ and $\ket{1}$ in a two-dimensional complex space, referred to as Hilbert space, with amplitudes $\alpha$ and $\beta$. Here the amplitudes are expressed in terms of Cartesian coordinates; but bear in mind that these coordinates can be complex numbers. 

\medskip
In a nutshell, consider the squared values of $\alpha$ and $\beta$ as probability values representing the likelihood of the measurement return 0 or 1.  For instance, let us assume the following: 

\begin{equation}
\ket{\Psi} = \alpha \ket{0} + \ket{1},  \ \ \text{where} \ \  \left| \alpha \right| = \frac{1}{2} \ \ \text{and} \ \ \left| \beta \right| = \frac{\sqrt{3}}{2}
\end{equation}

\begin{figure}[htbp]
\begin{center}\vspace{0.3cm}
\includegraphics[width=0.3\linewidth]{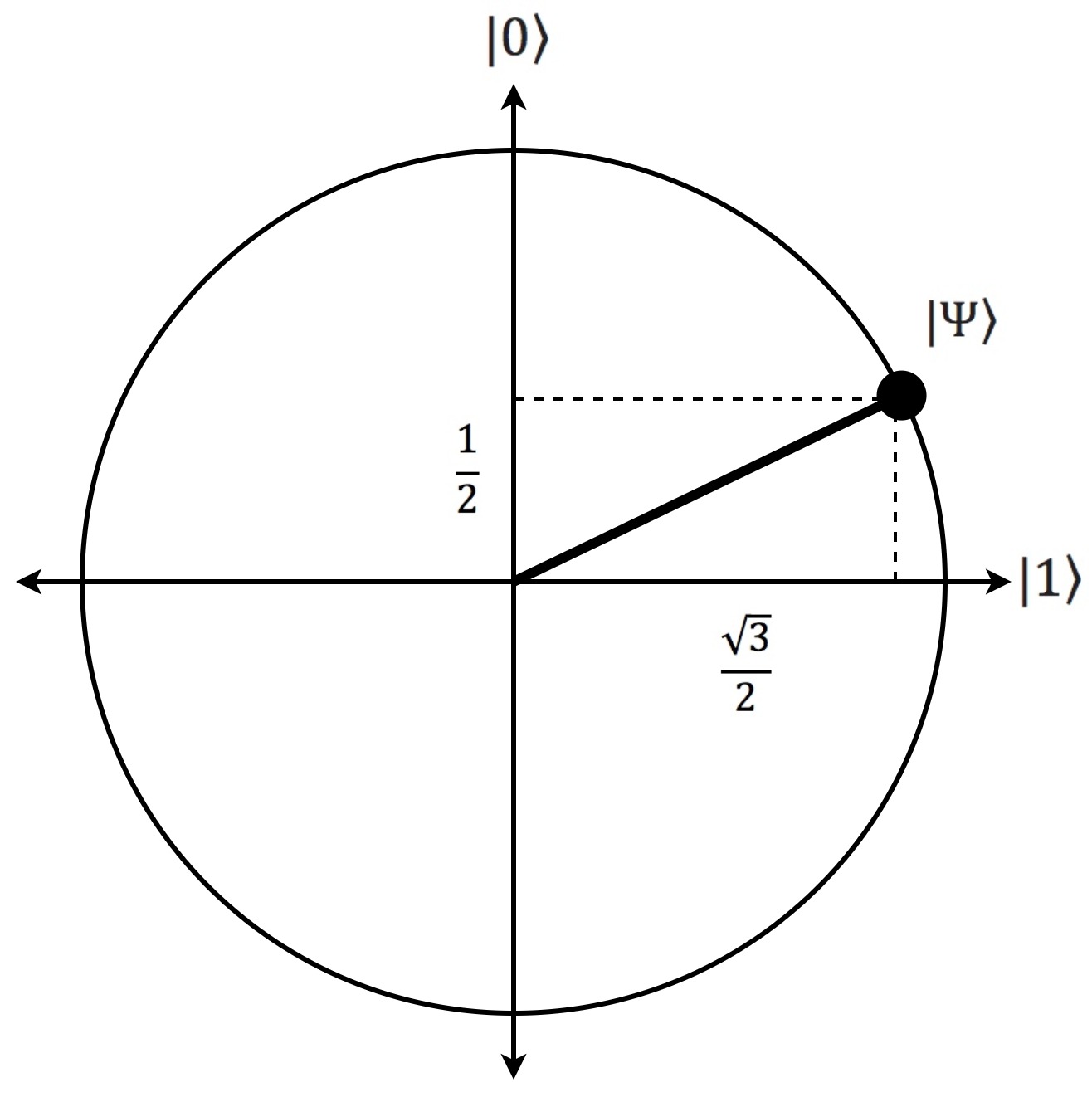}
\caption{An example of superposition, where the state vector has a 25\% chance of settling to $\ket{0}$ and a 75\% chance of settling to $\ket{1}$ after the measurement.}
\label{fig:radial-superp}
\end{center}
\end{figure}

\medskip
In this case, $\left| \alpha \right|^2=0.25$ and $\left| \beta \right|^2=0.75$. This means that the measurement of the qubit has a 25\% chance of returning 0 and a 75\% chance of returning 1 (Figure (\ref{fig:radial-superp}).

\begin{figure}[htbp]
\begin{center}\vspace{0.3cm}
\includegraphics[width=0.6\linewidth]{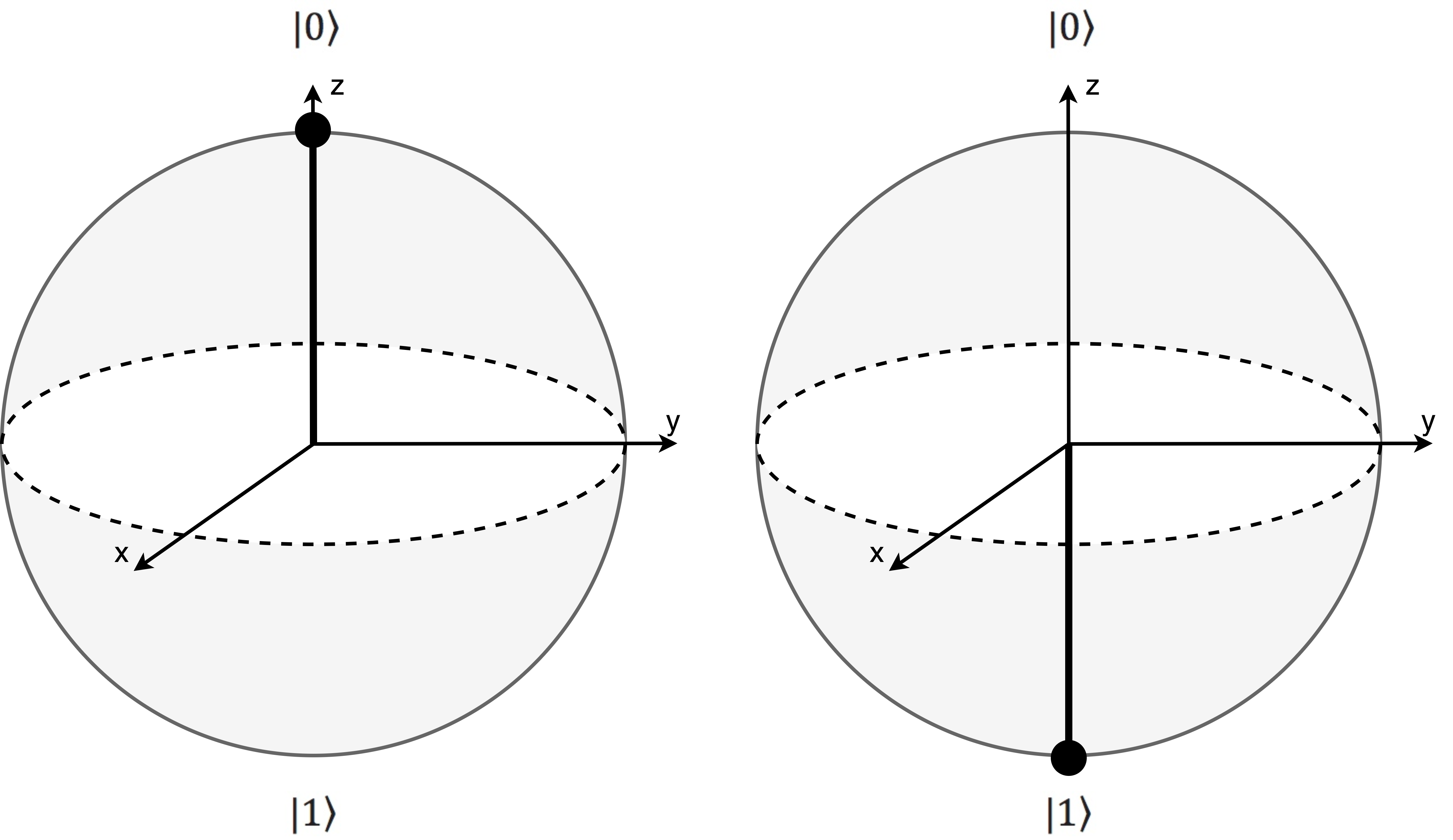}
\caption{The $\mathrm{\mathbf{X}}$ gate rotates the state vector (pointing upwards on the figure on the left) by 180 degrees around the x-axis (pointing downwards on the figure on the right).}
\label{fig:Bloch-X}
\end{center}
\end{figure}

\medskip
Quantum computers are programmed using sequences of commands, or quantum gates, that act on qubits. For instance, the ‘not gate’, performs a rotation of 180 degrees around the x-axis. Hence this gate is often referred to as the ‘$\mathrm{\mathbf{X}}$ gate’ (Figure \ref{fig:Bloch-X}).  A more generic rotation $\mathrm{\mathbf{Rx}}(\theta)$ gate is typically available for quantum programming, where the angle for the rotation around the x-axis is specified. Therefore, $\mathrm{\mathbf{Rx}}(180)$ applied to $\ket{0}$ or $\ket{1}$  is equivalent to applying $\mathrm{\mathbf{X}}$ to $\ket{0}$  or $\ket{1}$.  Similarly, also there are $\mathrm{\mathbf{Rz}}(\varphi)$ and $\mathrm{\mathbf{Ry}}(\theta)$ gates for rotations on the z-axis and y-axis, respectively.  As $\mathrm{\mathbf{Rz}}(180)$ is widely used in quantum computing, various programming tools provide the gate $\mathrm{\mathbf{Z}}$ to do this. An even more generic gate is typically available, which is a unitary rotation gate, with 3 Euler angles: $\mathrm{\mathbf{U(\theta, \varphi, \lambda)}}$. 

\medskip
In essence, all quantum gates perform rotations, which change the amplitude distribution of the system. And in fact, any qubit rotation can be specified in terms of $\mathrm{\mathbf{U(\theta, \varphi, \lambda)}}$; for instance $\mathrm{\mathbf{Rx}}(\theta) = \mathrm{\mathbf{U}}(\theta, - \frac{\pi}{2}, \frac{\pi}{2}$).

\begin{figure}[htbp]
\begin{center}\vspace{0.3cm}
\includegraphics[width=0.3\linewidth]{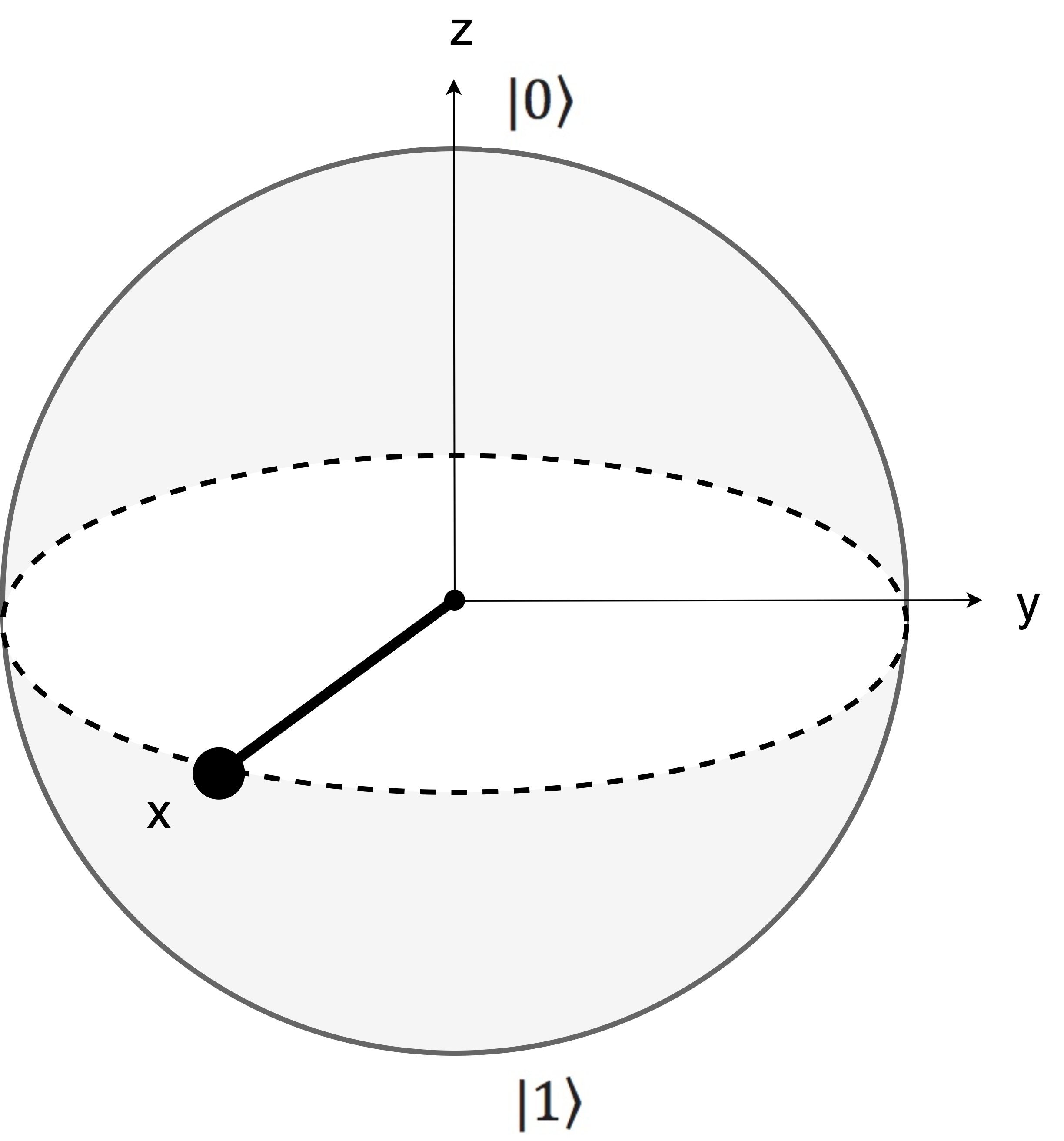}
\caption{The Hadamard gate puts the qubit into a superposition state halfway two opposing poles. }
\label{fig:Bloch-H}
\end{center}
\end{figure}

\medskip
An important gate for quantum computing is the Hadamard gate (referred to as the ‘$\mathrm{\mathbf{H}}$ gate’). It puts the qubit into a balanced superposition state (pointing to $\ket{+}$) consisting of an equal-weighted combination of two opposing states:$\left| \alpha \right|^2=0.5$ and   $\left| \beta \right|^2=0.5$. (Figure \ref{fig:Bloch-H}). For other gates, please consult the references given above.

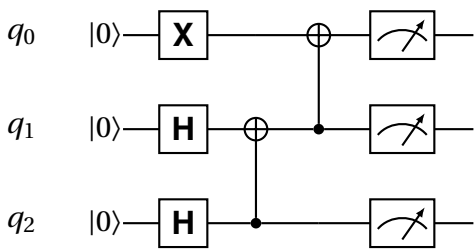
\begin{figure}[htbp]
\begin{center}\vspace{0.3cm}
    \begin{tikzpicture}
        \node[scale=1.0] 
        {
            \begin{quantikz}
                \lstick{$q_0$} &  \ket{0} & \gate{\textbf{X}}  & \qw        & \targ{}    &  \meter{}  & \qw \\
                \lstick{$q_1$} &  \ket{0} & \gate{\textbf{H}}  & \targ{}    & \ctrl{-1}  &  \meter{} & \qw \\
                \lstick{$q_2$} &  \ket{0} & \gate{\textbf{H}}  & \ctrl{-1}  & \qw        & \meter{} & \qw
            \end{quantikz}
        };
    \end{tikzpicture}
\end{center}
\caption{A quantum program depicted as a circuit of quantum gates. The squares with dials represent measurements, which are saved on classic registers.}
\label{fig:gen-circ}
\end{figure}

\medskip
A quantum program is often depicted as a circuit diagram of quantum gates, showing sequences of gate operations on the qubits (Figure \ref{fig:gen-circ}). Qubits typically start at $\ket{0}$ and then a sequence of gates are applied. Then, the qubits are read and the results are stored in standard digital memory, which are accessible for further handling. Normally a quantum computer works alongside a classical computer, which in effect acts as the interface between the user and the quantum machine. The classical machine enables the user to handle the measurements for practical applications.

\medskip
Quantum computation gets really interesting with gates that operate on multiple qubits, such as the conditional $\mathrm{\mathbf{X}}$ gate, or ‘$\mathrm{\mathbf{CX}}$ gate’. The $\mathrm{\mathbf{CX}}$ gate puts two qubits in entanglement. 

\begin{figure}[htbp]
\begin{center}\vspace{0.3cm}
    \begin{tikzpicture}
        \node[scale=1.0] 
        {
            \begin{quantikz}
                \lstick{$q_0$} &  \ket{0} & \ctrl{1}    & \qw \\
                \lstick{$q_1$} &  \ket{0} & \targ{}     & \qw
            \end{quantikz}
        };
    \end{tikzpicture}
\end{center}
\caption{The $\mathrm{\mathbf{CX}}$ gate creates a dependency of the state of one qubit with the state of another. In this case, $q_1$ will be flipped only if $q_0$ is $\ket{1}$.}
\label{fig:cnot}
\end{figure}
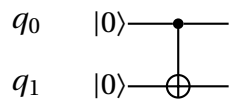

\medskip
Entanglement establishes a curious correlation between qubits. In practice, the $\mathrm{\mathbf{CX}}$ gate applies an $\mathrm{\mathbf{X}}$ gate on a qubit only if the state of another qubit is $\ket{1}$. Thus, the $\mathrm{\mathbf{CX}}$ gate establishes a dependency of the state of one qubit with the value of another (Figure \ref{fig:cnot}). In practice, any quantum gate can be made conditional and entanglement can take place between more than two qubits.

\medskip
The Bloch sphere is useful to visualize what happens with a single qubit, but it is not suitable for multiple qubits, in particular when they are entangled. Entangled qubits can no longer be thought of as independent units. They become one quantum entity described by a state vector of its own right on a hypersphere. A hypersphere is an extension of the Bloch sphere to $2^n$ complex dimensions, where $n$ is the number of qubits. Quantum gates perform rotations of a state vector to a new position on this hypersphere. Thus, it is virtually impossible to visualize a system with multiple qubits. Hence, from now on we shall use mathematics to represent quantum systems.

\medskip
The notation used above to represent quantum states ($\ket{\Psi}$, $\ket{0}$, $\ket{1}$), is referred to as Dirac notation, which provides an abbreviated way to represent a vector. For instance, $\ket{0}$ and $\ket{1}$), represent the following vectors, respectively:

\begin{align}
    	\ket{0} = \begin{bmatrix} 1 \\ 0 \end{bmatrix}			
 	\ \ \ \text{and} \ \ \
	\ket{1} = \begin{bmatrix} 0 \\ 1  \end{bmatrix}
\end{align}

And quantum gates are represented as matrices. For instance, the simples gate of them all is the, ‘$\mathrm{\mathbf{I}}$ gate’, or Identity gate, which is represented as an identity matrix:

\begin{align}
    	\mathrm{\mathbf{I}} = \begin{bmatrix} 
	1 & 0 \\ 
	0 & 1
	\end{bmatrix}			
\end{align}

The $\mathrm{\mathbf{I}}$ gate does not alter the state of a qubit. Thus, the application of an $\mathrm{\mathbf{I}}$ gate to $\ket{0}$ looks like this:

\begin{align}
    	\mathrm{\mathbf{I}}(\ket{0}) = \begin{bmatrix} 
	1 & 0 \\ 
	0 & 1
	\end{bmatrix}	
	\times
	\begin{bmatrix} 1 \\ 0 \end{bmatrix}	
	=
	\begin{bmatrix} 1 \\ 0  \end{bmatrix}	
	=	
	\ket{0} 
\end{align}

In contrast, the $\mathrm{\mathbf{X}}$ gate, which flips the state of a qubit is represented as:

\begin{align}
    	\mathrm{\mathbf{X}} = \begin{bmatrix} 
	0 & 1 \\ 
	1 & 0
	\end{bmatrix}			
\end{align}

\medskip
Therefore, quantum gate operations are represented mathematically as matrix operations; e.g., multiplication of a matrix (gate) by a vector (qubit state). Thus, the application of an $\mathrm{\mathbf{X}}$ gate to $\ket{0}$ looks like this:

\begin{align}
    	\mathrm{\mathbf{X}}(\ket{0}) = \begin{bmatrix} 
	0 & 1 \\ 
	1 & 0
	\end{bmatrix}	
	\times
	\begin{bmatrix} 1 \\ 0 \end{bmatrix}	
	=
	\begin{bmatrix} 0 \\ 1  \end{bmatrix}	
	=	
	\ket{1} 
\end{align}

\medskip
Conversely, the application of an $\mathrm{\mathbf{X}}$ gate to $\ket{1}$ would therefore is written as follows:

\begin{align}
    	\mathrm{\mathbf{X}}(\ket{1}) = \begin{bmatrix} 
	0 & 1 \\ 
	1 & 0
	\end{bmatrix}	
	\times
	\begin{bmatrix} 0 \\ 1 \end{bmatrix}	
	=
	\begin{bmatrix} 1 \\ 0  \end{bmatrix}	
	=	
	\ket{0} 
\end{align}

\medskip
The Hadamard gate has the matrix:

\begin{align}
    	\mathrm{\mathbf{H}} = \begin{bmatrix} 
	\frac{1}{\sqrt{2}} & \frac{1}{\sqrt{2}} \\ \\
	\frac{1}{\sqrt{2}} & - \frac{1}{\sqrt{2}}
	\end{bmatrix}	
	=
	\frac{1}{\sqrt{2}}
	\begin{bmatrix} 
	1 & 1 \\
	1 & - 1
	\end{bmatrix}		
\end{align}

\medskip
As we have seen earlier, the application of the H gate to a qubit pointing to $\ket{0}$ puts it in superposition, right at the equator of the Bloch sphere. This is notated as follows:

\begin{align}
    	\mathrm{\mathbf{H}}(\ket{0}) = \frac{1}{\sqrt{2}}(\ket{0}+\ket{1})
\end{align}

\medskip
As applied to $\ket{1}$, it also puts it in superposition, but pointing to the opposite direction of the superposition shown above:

\begin{align}
    	\mathrm{\mathbf{H}}(\ket{1}) = \frac{1}{\sqrt{2}}(\ket{0}-\ket{1})
\end{align}

\medskip
In the preceding equations, the result of $\mathrm{\mathbf{H}}(\ket{0})$ and $\mathrm{\mathbf{H}}(\ket{1})$)  could written as $\ket{+}$ and $\ket{-}$, respectively. In a circuit, we could subsequently apply another gate to $\ket{+}$ to $\ket{-}$, and so on; e.g. $\mathrm{\mathbf{X}}(\ket{+}) = \ket{+}$.  

\medskip
The Hadamard gate is often used to change the so-called \textit{computational basis} of the qubit.  The z-axis $\ket{0}$ and $\ket{1}$ form the standard basis. The x-axis $\ket{+}$ and $\ket{-}$ form the so-called \textit{conjugate basis}. As we saw earlier, the application of $\mathrm{\mathbf{X}}(\ket{+})$ would not have much effect if we measure the qubit in the standard basis: it would still probabilistically return 0 or 1. However, it would be different if we were to measure it in the conjugate basis; it would deterministically return the value on the opposite side where the vector is aiming to. 

\medskip
Another commonly used basis is the circular basis (y-axis). A more detailed explanation of different bases and their significance to computation and measurement can be found in \cite{Bernhardt2019}. What is important to keep in mind is that changing the basis on which a quantum state is expressed, corresponds to changing the kind of measurement we perform, and so, naturally, it also changes the probabilities of measurement outcomes.

\medskip
Quantum processing with multiple qubits is represented by means of tensor vectors. A tensor vector is the result of the tensor product, represented by the symbol $\bigotimes$ of two or more vectors. A system of two qubits looks like this $\ket{0} \otimes \ket{0}$, but it is normally abbreviated to $\ket{00}$. It is useful to study the expanded form of the tensor product to follow how it works:

\begin{align}
    	\ket{00} 
	=  
	\ket{0} \otimes \ket{0}
	=
	\begin{bmatrix} 1 \\ 0 \end{bmatrix} \otimes \begin{bmatrix} 1 \\ 0 \end{bmatrix}
	=
	\begin{bmatrix} 
	1 & \times & 1 \\ 
	1 & \times & 0 \\
	0 & \times & 1 \\
	0 & \times & 0 
	\end{bmatrix}	
	=
	\begin{bmatrix} 1 \\ 0 \\ 0 \\ 0  \end{bmatrix}	
\end{align}

\medskip
Similarly, the other 3 possible states of a 2-qubits system are as follows:

\begin{align}
    	\ket{01} 
	=  
	\ket{0} \otimes \ket{1}
	=
	\begin{bmatrix} 1 \\ 0 \end{bmatrix} \otimes \begin{bmatrix} 0 \\ 1 \end{bmatrix}
	=
	\begin{bmatrix} 
	1 & \times & 0 \\ 
	1 & \times & 1 \\
	0 & \times & 0 \\
	0 & \times & 1 
	\end{bmatrix}	
	=
	\begin{bmatrix} 0 \\ 1 \\ 0 \\ 0  \end{bmatrix}	
\end{align}

\begin{align}
    	\ket{10} 
	=  
	\ket{1} \otimes \ket{0}
	=
	\begin{bmatrix} 0 \\ 1 \end{bmatrix} \otimes \begin{bmatrix} 1 \\ 0 \end{bmatrix}
	=
	\begin{bmatrix} 
	0 & \times & 1 \\ 
	0 & \times & 0 \\
	1 & \times & 1 \\
	1 & \times & 0 
	\end{bmatrix}	
	=
	\begin{bmatrix} 0 \\ 0 \\ 1 \\ 0  \end{bmatrix}	
\end{align}

\begin{align}
    	\ket{11} 
	=  
	\ket{1} \otimes \ket{1}
	=
	\begin{bmatrix} 0 \\ 1 \end{bmatrix} \otimes \begin{bmatrix} 0 \\ 1 \end{bmatrix}
	=
	\begin{bmatrix} 
	0 & \times & 0 \\ 
	0 & \times & 1 \\
	1 & \times & 0 \\
	1 & \times & 1 
	\end{bmatrix}	
	=
	\begin{bmatrix} 0 \\ 0 \\ 0 \\ 1  \end{bmatrix}	
\end{align}

\medskip
We are now in a position to explain how the $\mathrm{\mathbf{CX}}$ gate works in more detail. This gate is defined by the matrix:

\begin{align}
    	\mathrm{\mathbf{CX}} = \begin{bmatrix} 
	1 & 0 & 0 & 0 \\ 
	0 & 1 & 0 & 0 \\
	0 & 0 & 0 & 1 \\
	0 & 0 & 1 & 0 
	\end{bmatrix}			
\end{align}

\medskip
The application of $\mathrm{\mathbf{CX}}$ to $\ket{10}$ is represented as:

\begin{align}
    	\mathrm{\mathbf{CX}}\ket{10} = 
	\begin{bmatrix} 
	1 & 0 & 0 & 0 \\ 
	0 & 1 & 0 & 0 \\
	0 & 0 & 0 & 1 \\
	0 & 0 & 1 & 0 
	\end{bmatrix}	
	\times
	\begin{bmatrix} 0 \\ 0 \\ 1\\ 0 \end{bmatrix} 	
	=
	\begin{bmatrix} 0 \\ 0 \\ 0\\ 1 \end{bmatrix}
	=
	\begin{bmatrix} 0 \\ 1 \end{bmatrix} \otimes \begin{bmatrix} 0 \\ 1 \end{bmatrix}
	=
	\ket{1} \otimes \ket{1} 
	=
	\ket{11}
\end{align}

\medskip
Table \ref{tab:cnot} shows the resulting quantum states of $\mathrm{\mathbf{CX}}$ gate operations, where the first qubit flips only if the second qubit is 1. Figure \ref{fig:cnot} illustrates how the $\mathrm{\mathbf{CX}}$ gate is represented in a circuit diagram. Note that in quantum computing qubit strings are often enumerated from the right end of the string to the left: $\dots \ket{q_2} \otimes \ket{q_1} \otimes \ket{q_0}$. (This is the convention adopted for the examples below.)
\begin{table}[]
\centering
\begin{tabular}{|c|c|}
\hline
\textbf{{\footnotesize Input}} & \textbf{{\footnotesize Result}} \\ \hline
$\ket{00}$  & $\ket{00}$ \\ \hline
$\ket{01}$  & $\ket{11}$ \\ \hline
$\ket{10}$  & $\ket{10}$ \\ \hline
$\ket{11}$  & $\ket{01}$ \\ \hline
\end{tabular}
\caption{\footnotesize{The $\mathrm{\mathbf{CX}}$ gate table, where $q_1$ changes only if $q_0$ is $\ket{1}$.} Note, by convention $\ket{q_1 q_0}$.}
\label{tab:cnot}
\end{table}

\begin{figure}[htbp]
\begin{center}\vspace{0.3cm}
    \begin{tikzpicture}
        \node[scale=1.0] 
        {
            \begin{quantikz}
                \lstick{$q_0$} &  \ket{0} & \ctrl{2}    & \qw \\
                \lstick{$q_1$} &  \ket{0} & \ctrl{1}     & \qw \\
                \lstick{$q_2$} &  \ket{0} & \targ{}     & \qw
            \end{quantikz}
        };
    \end{tikzpicture}
\end{center}
\caption{The Toffoli gate creates a dependency of the state of one qubit with the state of two others.}
\label{fig:Toffoli}
\end{figure}
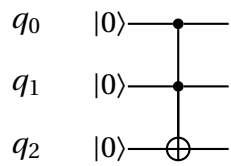

\medskip
Another useful conditional gate, which appears on a number of quantum algorithms, is the $\mathrm{\mathbf{CCX}}$ gate, also known as the Toffoli gate, involving three qubits (Figure \ref{fig:Toffoli}). Table 2 shows resulting quantum states of the Toffoli gate: qubit $q_2$ flips only if both $q_1$ and $q_0$ are $\ket{1}$.

\begin{table}[]
\centering
\begin{tabular}{|c|c|}
\hline
\textbf{{\footnotesize Input}} & \textbf{{\footnotesize Result}} \\ \hline
$\ket{000}$  & $\ket{000}$ \\ \hline
$\ket{001}$  & $\ket{001}$ \\ \hline
$\ket{010}$  & $\ket{010}$ \\ \hline
$\ket{011}$  & $\ket{111}$ \\ \hline
$\ket{100}$  & $\ket{100}$ \\ \hline
$\ket{101}$  & $\ket{101}$ \\ \hline
$\ket{110}$  & $\ket{110}$ \\ \hline
$\ket{111}$  & $\ket{011}$ \\ \hline
\end{tabular}
\caption{\footnotesize{Toffoli gate table. Note, by convention $\ket{q_2 q_1 q_0}$}}
\label{tab:Toffoli}
\end{table}

\medskip
The equation for describing a 2-qubits system $\ket{q_1}$ and $\ket{q_0}$ combines two state vectors $\ket{\Psi}$ and $\ket{\Phi}$ as follows. Consider:

\begin{equation}
\begin{split}
    	\ket{\Psi} = \alpha_1\ket{0} + \alpha_2\ket{1} \ \text{for}  \ q_0 \\ \\
	\ket{\Phi} = \beta_1\ket{0} + \beta_2\ket{1}  \ \text{for}  \ q_1
\end{split}
\end{equation}

\medskip
Then:

\begin{align}
    	\ket{\Psi} \otimes \ket{\Phi}  = \alpha_0\beta_0\ket{00} + \alpha_0\beta_1\ket{01} + \alpha_1\beta_0\ket{10} + \alpha_1\beta_1\ket{11}
\end{align}

\medskip
The above represents a new quantum state with four amplitude coefficients, which can be written as:

\begin{align}
    	\ket{D}  = \delta_0\ket{00} + \delta_1\ket{01} + \delta_2\ket{10} + \delta_3\ket{11}
\end{align}

\medskip
Consider this equation:

\begin{align}
    	\ket{\Psi}  = \frac{1}{4}\ket{00} + \frac{1}{4}\ket{01} + \frac{1}{4}\ket{10} + \frac{1}{4}\ket{11}
\end{align}

\medskip
The above is saying that each of the four quantum states have equal probability of 25\% each of being returned.

\medskip
Now, it should be straightforward to work out how to describe quantum systems with more qubits. For instance, a system with four qubits looks like this:

\begin{equation}
	\begin{split}
    	\ket{B}  = \beta_0\ket{0000} + \beta_1\ket{0001} + \beta_2\ket{0010} + \beta_3\ket{0011} + \\
	\beta_4\ket{0100} + \beta_5\ket{0101} + \beta_6\ket{0110} + \beta_7\ket{0111} +  \\ 
	 \beta_8\ket{1000} + \beta_9\ket{1001} + \beta_{10}\ket{1010} + \beta_{11}\ket{1011} + \\ 
	 \beta_{12}\ket{1100} + \beta_{13}\ket{1101} + \beta_{14}\ket{1110} + \beta_{15}\ket{1111}
	\end{split}
\end{equation}

\medskip
A linear increase of the number of qubits extends the capacity of representing information on a quantum computer exponentially. With qubits in superposition, a quantum computer can handle all possible values of some input data simultaneously. This endows the machine with massive parallelism. However, we do not have access to the information until the qubits are measured. 

\medskip
Quantum algorithms require a different way of thinking than the way one normally approaches programming; for instance, it is not possible to store quantum states on a working memory for accessing later in the algorithm. This is due to the so-called non-cloning principle of quantum physics: it is impossible to make a copy of a quantum system. It is possible, however, to move the state of a set of qubits to another set of qubits, but in effect this deletes the information from the original qubits. To program a quantum computer requires manipulations of qubits so that the states that correspond to the desired outcome have a much higher probability of being measured than all the other possibilities. 

\medskip
Decoherence is problematic because it poses limitations on the number of successive gates we can use in a circuit (a.k.a. the circuit depth). The higher the number of gates sequenced one after the other (i.e., circuit depth) and the number of qubits used (i.e., circuit width), the higher the likelihood of decoherence to occur. At the time of writing, quantum processors struggle to maintain a handful of qubits coherent for more than a dozen successive gates involving superposition and entanglement. Effectively, most of the circuits introduced in this chapter are far deeper than the critical depth for which existing quantum devices can maintain coherence.

\medskip
One way to mitigate errors is to run the algorithms many times and then select the result that appeared most. Additional post processing on the measurement outcomes that tries to undo the effect of the noise by solving an inverse problem can also be carried out. The development of sophisticated error correction methods is an important research challenge.

%
%


\end{document}